\documentclass[line]{aastex631}

	\usepackage{graphicx}	
	\usepackage{amsmath}
	\usepackage{amssymb}	
	\usepackage{longtable}
	\usepackage{threeparttable}
	\usepackage{multirow}
	\usepackage{booktabs}
  \usepackage{natbib}
	\usepackage{array}
    \usepackage[figuresright]{rotating}
    \usepackage[T1]{fontenc}

	\graphicspath{{./}{figures/}}

\begin{document}
	  \title{Using Asteroseismology to Calibrate the Physical Parameters of \\
      Confirmed Exoplanets and Their Evolved Host Stars}

  \correspondingauthor{Sheng-Bang Qian}
  \email{qsb@ynao.ac.cn}

  \author[0000-0002-8276-5634]{Wen-Xu Lin}
  \affiliation{Yunnan Observatories, Chinese Academy of Sciences, Kunming 650216, People's Republic of China}
  \affiliation{Department of Astronomy, School of Physics and Astronomy, Yunnan University, Kunming 650091, P. R. China}
  \affiliation{Key Laboratory of Astroparticle Physics of Yunnan Province, Yunnan University, Kunming 650091, P. R. China}
  \affiliation{University of Chinese Academy of Sciences, No.1 Yanqihu East Rd, Huairou District, Beijing, 101408, People's Republic of China}

  \author[0000-0002-5995-0794]{Sheng-Bang Qian}
  \affiliation{Department of Astronomy, School of Physics and Astronomy, Yunnan University, Kunming 650091, P. R. China}
  \affiliation{Key Laboratory of Astroparticle Physics of Yunnan Province, Yunnan University, Kunming 650091, P. R. China}

  \author[0000-0002-0796-7009]{Li-Ying Zhu}
  \affiliation{Yunnan Observatories, Chinese Academy of Sciences, Kunming 650216, People's Republic of China}
  \affiliation{University of Chinese Academy of Sciences, No.1 Yanqihu East Rd, Huairou District, Beijing, 101408, People's Republic of China}

  \author[0000-0001-9346-9876]{Wen-Ping Liao}
  \affiliation{Yunnan Observatories, Chinese Academy of Sciences, Kunming 650216, People's Republic of China}
  \affiliation{University of Chinese Academy of Sciences, No.1 Yanqihu East Rd, Huairou District, Beijing, 101408, People's Republic of China}

  \author[0000-0002-0285-6051]{Fu-Xing Li}
  \affiliation{Department of Astronomy, School of Physics and Astronomy, Yunnan University, Kunming 650091, P. R. China}
  \affiliation{Key Laboratory of Astroparticle Physics of Yunnan Province, Yunnan University, Kunming 650091, P. R. China}

  \begin{abstract}

    Asteroseismology offers a profound window into stellar interiors and has emerged as a pivotal technique in exoplanetary research. This study harnesses the Transiting Exoplanet Survey Satellite (TESS) observations to reveal, for the first time, the asteroseismic oscillations of four exoplanet-hosting stars. Through meticulous analysis, we extracted their asteroseismic signatures, enabling the precise determination of stellar masses, radii, luminosities, and surface gravities. These parameters exhibit markedly reduced uncertainties compared to those derived from spectroscopic methods. Crucially, the exoplanets orbiting these stars were initially identified via radial velocity measurements. The refinement of host stellar masses necessitates a corresponding adjustment in planetary characteristics. Employing asteroseismology, we recalibrated the exoplanets' minimum masses and semi-major axes—a novel approach in the field. For instance, the exoplanet HD 5608 b's minimum mass, denoted as $M\sin{i}$, was ascertained to be 1.421$\pm 0.091 M_{J}$ through the integration of asteroseismic and radial velocity data. Similarly, two planets within the 7 CMa system yielded $M\sin{i}$ values of 1.940$\pm 0.064 M_{J}$ and 0.912$\pm 0.067 M_{J}$, respectively. Two planets in HD 33844 system presented $M\sin{i}$ figures of 1.726$\pm 0.145 M_{J}$ and 1.541$\pm 0.182 M_{J}$, while the HIP 67851 system's planets registered $M\sin{i}$ at 1.243$\pm 0.139 M_{J}$ and a notably higher 5.387$\pm 0.699 M_{J}$. This investigation extends beyond mere parameter refinement; it underscores the synergy between asteroseismology and exoplanetology, yielding unprecedented precision in system metrics. Focusing on quartet of K-type giants in advanced evolutionary phases, our work positions these systems as invaluable astrophysical laboratories, offering insights into the potential trajectory of our own solar system's fate.

  \end{abstract}


  \section{Introduction} \label{sec:intro}

  To study the internal structure of stars and determine their evolutionary stages, asteroseismology has become the most powerful tool, opening a new window for the field of astronomy. Thanks to the high-precision, long-duration observations of space telescopes in recent years, such as ESA’s \textbf{Co}nvection, \textbf{Ro}tation and planetary \textbf{T}ransits, $CoRoT$\citep{2009IAUS..253...71B}, NASA’s $Kepler$/$K2$\citep{2010Sci...327..977B,2014PASP..126..398H} and \textbf{T}ransiting \textbf{E}xoplanet \textbf{S}urvey \textbf{S}atellite, $TESS$\citep{ricker2015transiting}, a large amount of light curve data has greatly advanced the research of asteroseismology. Therefore, we have the opportunity to study solar-like stars and red giants more deeply, which exhibit solar-like oscillations driven by convection\citep{chaplin2013asteroseismology}.

  Solar-like oscillations have two different types of standing wave modes driven by different physical effects, pressure modes (p modes) with gradient pressure as a restoring force, and internal gravity waves (g modes) associated with the effects of buoyancy\citep{chaplin2013asteroseismology}. Solar-like oscillations excited by surface convection also exist in red giants\citep{frandsen2002detection}, which means that we have the opportunity to explore the rich structural information inside red giants. 
  In asteroseismic studies of evolved stars, two global seismic parameters are important: the maximum amplitude frequency, $\nu_{max}$, and the mean maximum separation, $\Delta \nu$, which reflect global stellar properties and, when combined with estimates of stellar surface temperature, allow the calculation of the radius and mass of the star\citep{kjeldsen1994amplitudes,miglio2009probing,bedding2010solar}. And through precise spectroscopic and asteroseismic measurements, it is demonstrated that the two independent methods of obtaining the physical parameters of the star are consistent \citep{coelho2024seismic}.

  As asteroseismology has become a powerful tool for exploring the interiors of stars, it has gradually gained prominence in various fields of astronomy, especially exoplanet science \citep{campante2018asteroseismology}. The determination of exoplanet parameters with unparalleled precision is possible through astroseismology measurements of the exoplanet's host star \citep{huber2013fundamental, silva2015ages, lundkvist2016hot}. And in multiplanet systems, asteroseismology enables the discovery of misalignments between the rotation axis of the host star and the planetary disk, which in turn disrupts the inherent perception of planetary orbital migration \citep{huber2013stellar}. In the case of exoplanets discovered using the radial velocity method, the asteroseismic study of the host star also plays a crucial role in the evolution of the overall system \citep{campante2019tess}. Planets at very short distances from their host stars are easily noticeable, and when the host star is found to be a red cluster giant by asteroseismic analysis, then how the planet escaped being engulfed becomes the focus of research \citep{hon2023close}. As demonstrated above, the combination of asteroseismology and exoplanetology has the potential to significantly enhance research outcomes. This highlights the importance of interdisciplinary fields in astronomy.

  This work presents asteroseismic analyses of four host stars with confirmed exoplanets. The sources, HD 5608, 7 CMa, HD 33844, and HIP 67851, were cross-compared using TESS observations, revealing significant asteroseismic light variations. Utilized the $Lightkurve$ package \citep{2018ascl.soft12013L} to preprocess the light curve to obtain the power spectrum, then employed the $pySYD$ package \citep{2022JOSS....7.3331C} to acquire the observed global seismic parameters of these host stars, and used the $PBjam$$\footnote{https://github.com/grd349/PBjam}$ package to determine the observed radial, $\nu_{n,0}$, and quadrupolar, $\nu_{n,2}$, oscillation modes. Then applied the equations of the asteroseismic parameters with the physical parameters of stars to obtain physical parameters independent of the spectroscopic analysis. According to the calculation results of the physical parameters of the four host stars, the errors obtained from asteroseismology are smaller than those obtained from spectroscopy in other papers, but this does not mean that the parameters obtained from asteroseismology are more accurate. It has been tested
  theoretically and observationally, with a typical accuracy of
  $\sim$5$\%$ and $\sim$10$\%$–15$\%$ in radius and mass for red giants,
  respectively \citep{miglio2011asteroseismology,aguirre2012verifying,coelho2024seismic}. Based on this foundation and the radial velocity data obtained in the previous study, we recalibrated the exoplanet's minimum mass and orbital semi-major axis. We compiled all the calibrated parameters into a table for comparison with the previous study's results. Finally, this work's focus and significance are analyzed, with the hope of providing a basis for future exoplanet analysis and more in-depth analyses.

  \begin{figure}[ht!]

    \includegraphics[width=0.9\linewidth]{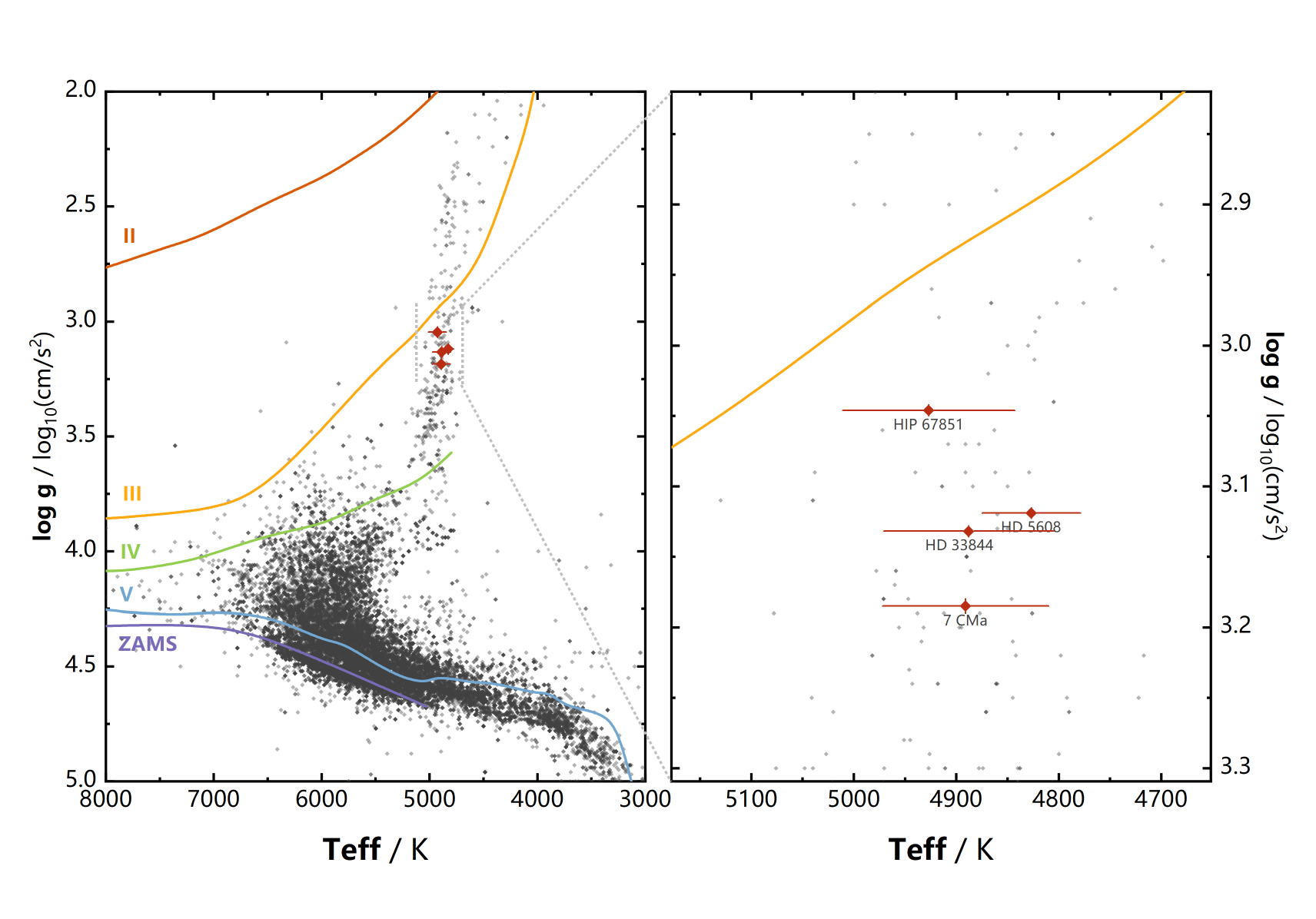}
    \centering
    \caption{Positions of the four host stars in the HR-gram. The horizontal coordinate is the equivalent temperature of the star and the vertical coordinate is the surface gravity of the star. The four red points with error bars are the four host stars involved in this work, and the gray points are the more than 5,000 other confirmed host stars with planets. It can be seen that all four host stars are between giant (III, orange line) and subgiant (IV, green line) \citep{straivzys1981fundamental}. In addition, the red lines mark the positions of supergiants (II), the blue lines mark the positions of dwarfs (V) \citep{straivzys1981fundamental}, and the purple lines mark the positions of zero-age main sequence stars (ZAMS) \citep{cox2000theoretical}.
    \label{fig:logg-teff}}

  \end{figure}

  \section{Seismology Analysis} \label{sec:Seismology Analysis}

  In the frequency spectra of red giants, including radial and non-radial modes\citep{de2009non}, whose frequencies $\nu_{n,l}$ and p modes of higher radial order, $n$, and lower angular degree, $l$, basically follow an asymptotic relationship\citep{vandakurov1967frequency,tassoul1980asymptotic,bedding2010solar},
  \begin{equation}\label{equation:one}
    \nu _{n,l } \simeq \Delta \nu (n +\frac{l}{2} +\epsilon ) - \delta \nu _{0,l} .
  \end{equation}
  The $\delta \nu _{0,l}$ is the small separation of the non-radial modes relative to the radial modes\citep{stello2013asteroseismic}.The $\epsilon$ is the offset from zero of the fundamental radial mode in units of the large separation, which is the $\Delta \nu$, the frequency shift of consecutive overtone modes of the same degree, is approximately equal to the inverse of the sound travel time across the star\citep{bedding2010solar,chaplin2013asteroseismology},
  \begin{equation}\label{equation:two}
    \Delta \nu = (2\int_{0}^{R} \frac{dr}{c} )^{-1} ,
  \end{equation}
  $c$ being the sound speed and $R$ the stellar radius. The average large separation $\left \langle \Delta \nu _{nl} \right \rangle$   is good approximation equal to $\Delta \nu$. And it's indicated that $\left \langle \Delta \nu _{nl} \right \rangle$ scales to very good approximation with the square root of the mean density. The frequency of maximum oscillations power, $\nu _{max}$, has a scaling relationship with the stellar surface gravity and effective temperature\citep{chaplin2013asteroseismology},
  \begin{equation}\label{equation:3}
    \left \langle \Delta \nu _{nl} \right \rangle \propto \left \langle \rho  \right \rangle ^{\frac{1}{2} } ,
  \end{equation}
  \begin{equation}\label{equation:4}
    \nu_{max}  \propto g \cdot T_{eff}^{-\frac{1}{2} } .
  \end{equation}

  With estimates of $\nu _{max}$ and $\Delta \nu$, together with independent estimates of $T_{eff}$, we can estimate the radius, mass, mean density and surface gravity of the star. In practical applications, precise measurements of the solar values of parameters are needed to scale the parameters proportionally\citep{chaplin2013asteroseismology}, 
  \begin{equation}\label{equation:5}
    (\frac{R}{R_{\odot } } ) \simeq (\frac{\nu_{max}}{\nu_{max,\odot} } )(\frac{\left \langle \Delta\nu_{nl} \right \rangle }{\left \langle \Delta\nu_{nl} \right \rangle _{\odot} } )^{-2} (\frac{T_{eff}}{T_{eff,\odot} })^{0.5},
  \end{equation}
  \begin{equation}\label{equation:6}
    (\frac{M}{M_{\odot } } ) \simeq (\frac{\nu_{max}}{\nu_{max,\odot} } )^{3}(\frac{\left \langle \Delta\nu_{nl} \right \rangle }{\left \langle \Delta\nu_{nl} \right \rangle _{\odot} } )^{-4} (\frac{T_{eff}}{T_{eff,\odot} })^{1.5},
  \end{equation}
  \begin{equation}\label{equation:7}
    (\frac{\rho }{\rho_{\odot } } ) \simeq (\frac{\left \langle \Delta\nu_{nl} \right \rangle }{\left \langle \Delta\nu_{nl} \right \rangle _{\odot} } )^{2} ,
  \end{equation}
  \begin{equation}\label{equation:8}
    (\frac{g}{g_{\odot } } ) \simeq (\frac{\nu_{max}}{\nu_{max,\odot} } ) (\frac{T_{eff}}{T_{eff,\odot} })^{0.5}.
  \end{equation}

  The solar effective temperature, $T _{\odot } $, is equal to 5772.0 $\pm$ 0.8 K \citep{prvsa2016nominal}.
  The solar $\nu _{max}$ is equal to 3141 $\pm$ 12 $\mu$Hz, and the solar $\Delta \nu$ is equal to 134.98 $\pm$ 0.04 $\mu$Hz, which is consistent with the value of $\left \langle \Delta\nu_{nl} \right \rangle _{\odot}$ \citep{andersen2019oscillations}.

  \subsection{Preprocessing of light curves}

  We used Python package $lightkurve$ \citep{2018ascl.soft12013L} to download  all available 2-minute cadence light curves of research targets from the Mikulski Archive for Space Telescope (MAST, https://mast.stsci.edu/portal/Mashup/Clients/Mast/Portal.html). These light
  curves were extracted and de-trended by the TESS Science Processing Operations Center (SPOC) pipeline \citep{jenkins2016tess}. We then used the $seismolog$ module in $lightkurve$ to do the asteroseismic analysis of the light curves in order to estimate reference values of $\nu _{max}$ and $\Delta \nu$, however, this is not a final result as no specific error is given, and this step is necessary to prepare for the next calculation in $PBjam$.

  \subsection{Used pySYD to obtain observed global seismic parameters}

  The $pySYD$ is a Python package for detecting solar-like osillations and measuring global asteroseismic parameters, include $\nu _{max}$ and $\Delta \nu$. It's an open-source tool and have been extensively tested against some closed-source tools \citep{2022JOSS....7.3331C}. Due to the well-developed software package, only the light curve and corresponding power spectrum are needed, along with a simple instruction, to automatically complete the analysis. The $pySYD$ package searches for localized power excess due to solar-like oscillations and then estimates its initial properties, and uses estimates to mask out that region in the power spectrum and implements an automated background fitting routine that characterizes amplitudes and characteristic time scales of various granulation process. Then derives global asteroseismic quantities from the background-corrected power spectrum and performs Monte-Carlo simulations by drawing from chi-squared distribution to estimate uncertainties. The seismic parameters output for each host star is individually showcased in the following text.

  \subsection{Recognizing oscillation modes with PBjam}

  $PBjam$ is a peakbagging Python package for modeling the oscillation spectra of solar-like oscillators. This involves two main parts: identifying a set of modes of interest, and accurately modeling those modes to measure their frequencies \citep{nielsen2021pbjam}. To accomplish the above, there are three main steps \citep{patil2022improving}:
  \begin{itemize}
      \item [1.]
      \texttt{KDE}: First, a kernel density estimate (KDE) of the prior P($\theta$) is computed using previously fit $\theta$ of 13,288 Kepler stars, where $\theta$ represents the set of parameters of a solar-like power spectrum model. Then, the KDE prior and the inputs to PBjam are used to estimate a starting point for next step.
      \item [2.] 
      \texttt{Asy$\underline{~}$peakbag}: The output of the KDE class is used to fit the asymptotic relation to the spectrum. This provides the most probable frequency intervals for the final stage of peakbagging.
      \item [3.] 
      \texttt{Peakbag}: The last step of the process is to loosen the majority of the parameterization utilized in Asy$\underline{~}$peakbag. This provides a relatively unconstrained estimation of the mode frequencies and their uncertainties. This approach enables PBjam to capture small variations in the mode frequencies that are not accounted for by the asymptotic relation.
  \end{itemize}

  \subsection{Recalculating the minimum mass and orbital semi-major axis of exoplanets}
  Asteroseismology is a reliable method for calculating stellar parameters, as it produces results consistent with other calculation methods \citep{coelho2024seismic}. However, when astroseismology measures physical parameters of a star that differ from spectroscopy, it can affect the parameters of the planets in the system. The mass of the star is critical since this work involves stars using the radial velocity method for exoplanet discovery and no transit signal has been detected yet. Mass function applicable to planetary systems can be derived based on Newton's laws \citep{wright2017radial}:
  \begin{equation}\label{equation:9}
    f \equiv \frac{PK^{3} (1 - e^{2} )^{\frac{3}{2} } }{2\pi G} = \frac{M_{planet}^{3} sin^{3}  i}{(M_{planet} +M_{star})^{2} } 
  \end{equation}
  Here, $K$ represents the (semi-)amplitude of the radial velocity variations, $e$ represents the eccentricity of the orbit, $i$ represents the inclination of the plane of the orbit with respect to the plane of the sky, $P$ represents the period of the orbit and $G$ is Newton’s constant. The planetary mass is actually much smaller than the stellar mass, so the planetary mass can be neglected in the right-hand side denominator in the above equation. For scale, it may be helpful to express the quantities in units that are more commonly understood and transform the equation:
  \begin{equation}\label{equation:10}
    K = 203.429 m/s (1-e^{2})^{-\frac{1}{2} }(P/day)^{-\frac{1}{3}} (M_{star} /M_{\odot })^{-\frac{2}{3} } (M_{planet}/M_{Jupiter})\sin{i} 
  \end{equation}

  Kepler's third law provides the following relationship of orbital semi-major axis, $a$:
  \begin{equation}\label{equation:11}
    a = (\frac{GM_{star} T^{2}}{4\pi^{2}} )^{\frac{1}{3} }
  \end{equation}

  Variations in stellar mass directly affect the minimum mass and orbital semi-major axis of exoplanets, two crucial parameters for studying planetary systems. Since this work did not involve acquiring new data of radial velocities, it would not affect other parameters in the system.

  \section{Calculation of physical parameters using asteroseismology}\label{sec:4 Host Stars}

  \begin{table}[h!]
    \centering
    \caption{Asteroseismic parameters of 4 host stars}
    \label{tab:4 star}
    \tabletypesize{\scriptsize}
    \tablewidth{0pt}
    \renewcommand{\arraystretch}{1.5}
    \begin{tabular}{ccccc}
      \hline
      \hline
      & HD 5608 & 7 CMa & HD 33844 & HIP 67851 \\
      \hline
      {$\nu _{max}$ [$\mu Hz$] } & 169.532$\pm 0.614$ & 182.926$\pm 1.559$ & 165.951$\pm 1.171$ & 134.433$\pm 0.638$ \\
      {$\Delta \nu$ [$\mu Hz$] } & 12.348$\pm 0.146$  & 13.662$\pm 0.033$  & 12.665$\pm 0.120$  & 10.985$\pm 0.027$  \\
      \hline
    \end{tabular}
  \end{table}

  \subsection{HD 5608}\label{HD 5608}
\begin{figure}[ht!]

  \includegraphics[width=0.6\linewidth]{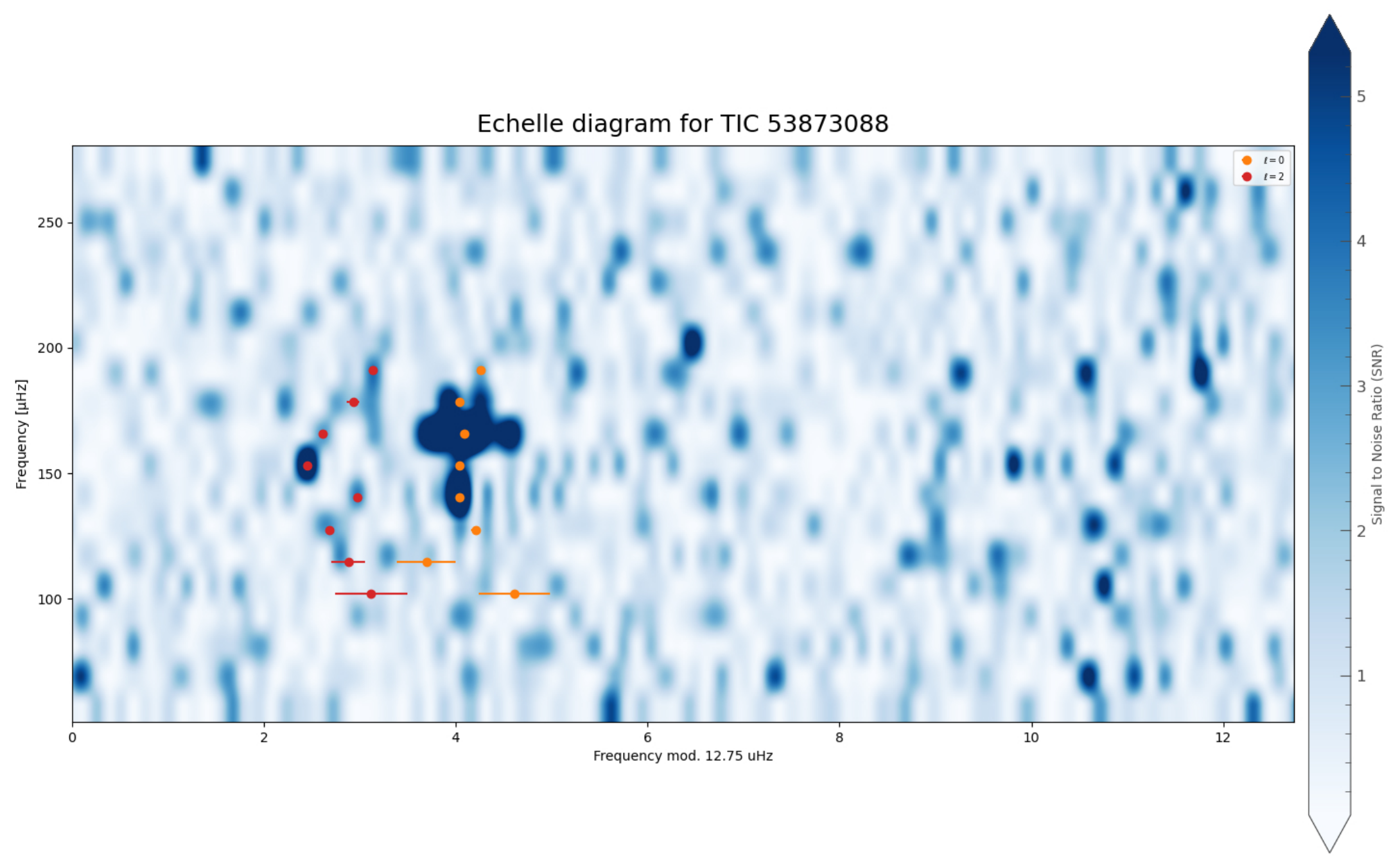}
  \centering
  \caption{HD 5608 Echelle diagram of smoothed SNR spectral estimate. Orange dots with error bars mark where the radial $l=0$ mode is located, and red dots with error bars mark where the quadrupole $l=2$ mode is located. These are estimated from Step 3. (\texttt{Peakbag}) of \texttt{PBjam}. The $\Delta \nu$ value here is not an actual observed value computed.
  \label{fig:HD 5608 1}}

  \centering
  \includegraphics[width=0.55\linewidth]{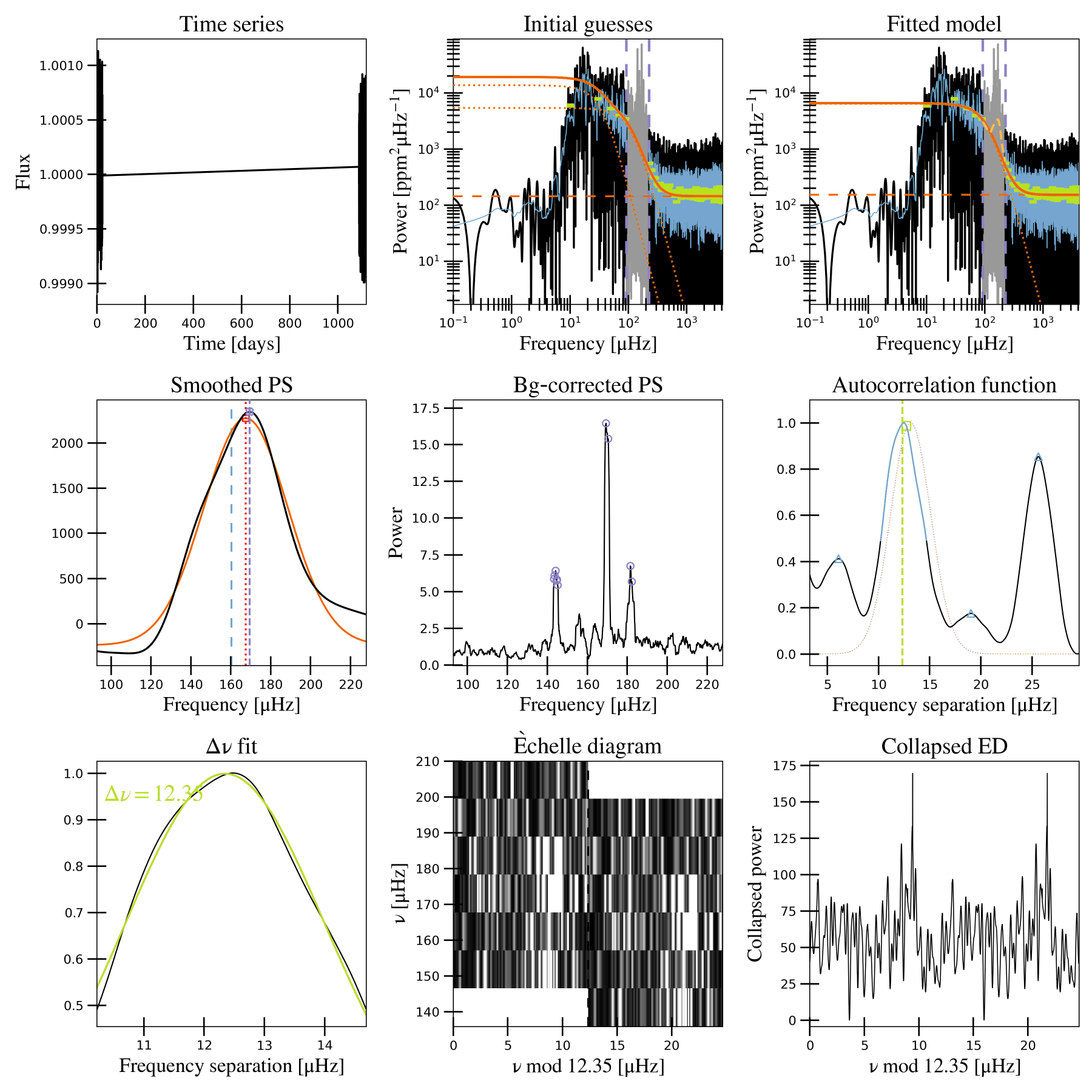}
  \caption{HD 5608 global asteroseismic parameter fitting. \textbf{Top left}: Original time series. \textbf{Top middle}: Original power spectrum (black), lightly smoothed power spectrum (blue), and binned power spectrum (green). Red-orange lines show initial guesses of the fit to the granulation background. The grey region is excluded from the background fit based on the numax estimate provided to the module. \textbf{Top right}: Same as top middle but now showing the best fit background model (red-orange) and a heavily smoothed version of the power spectrum (yellow). \textbf{Center left}: Background corrected, heavily smoothed power spectrum (black). The red-orange line shows a Gaussian fit to the data. \textbf{Center}: Lightly smoothed, background corrected power spectrum centered on $\nu _{max}$. \textbf{Center right}: Autocorrelation function of the data in the center panel. \textbf{Bottom left}: ACF peak extracted in the center right panel (black) and a Gaussian fit to that peak (green). The center of the Gaussian is the estimate of $\Delta \nu$. \textbf{Bottom middle}: Echelle diagram of the background corrected power spectrum using the measured $\Delta \nu$ value. \textbf{Bottom right}: Echelle diagram collapsed along the frequency direction.
  \label{fig:HD 5608 2}}

  \end{figure}

\begin{figure}[ht!]
  \includegraphics[width=0.58\linewidth]{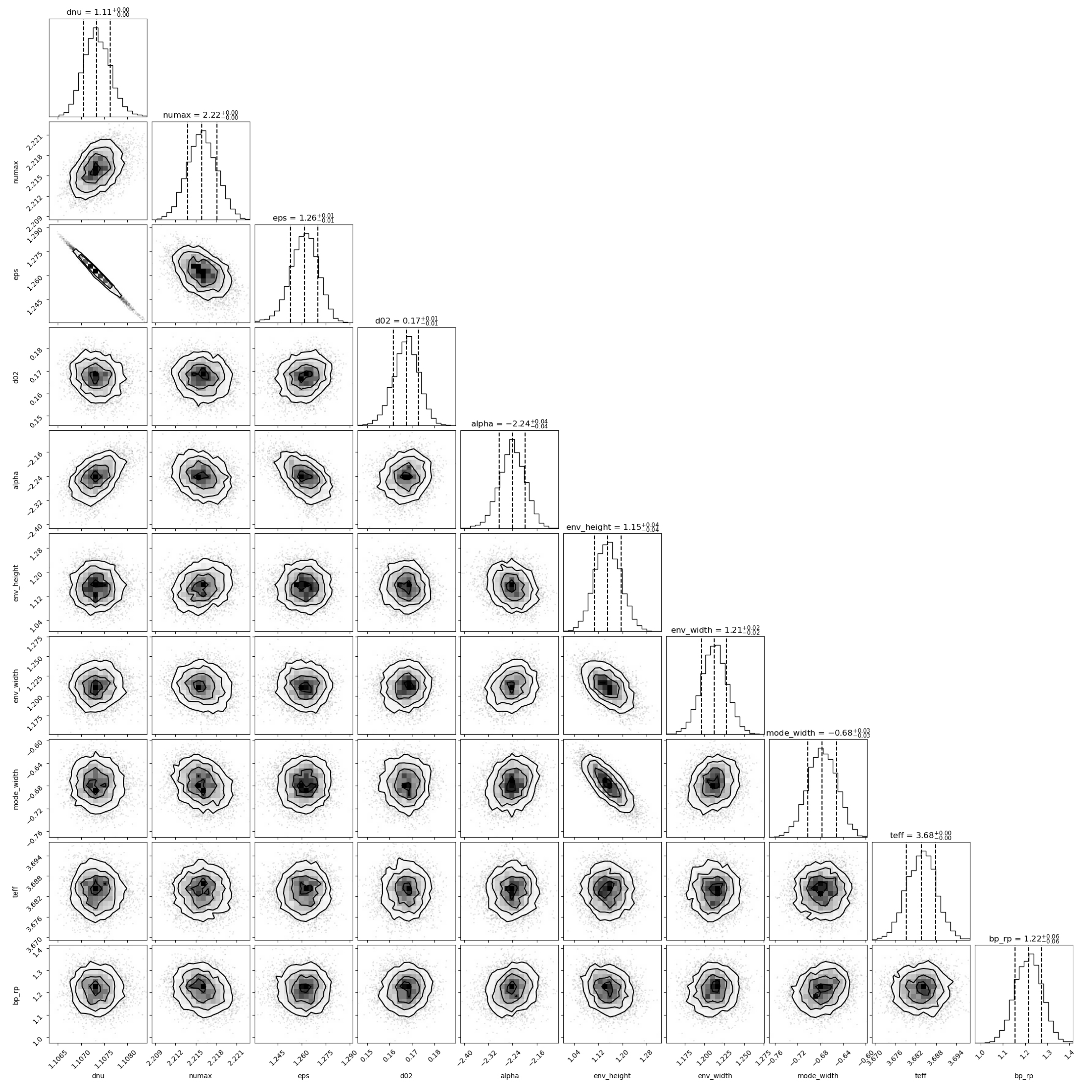}
  \centering
  \caption{HD 5608 corner diagram of asymptotic fit parameters.
  \label{fig:HD 5608 3}}
  \end{figure}

  HD 5608 (TIC 53873088) is a K0 IV giant with an exoplanet in its system, HD 5608 b, first discovered in 2012 \citep{sato2012substellar}. This target was observed by TESS in October 2019 and October 2022 as Sector 17 and Sector 57, respectively. The light curve of HD 5608 was de-trended using $lightkurve$. The $pySYD$ package was utilized to calculate $\Delta \nu$ and $\nu _{max}$. Next, by adding the effective temperature of the star as given in articles \footnote{Articles shown in Table \ref{tab:HD 5608}, and the effective temperatures in the different articles are given in the second column of the table.} and analyzing the spectral data using $PBjam$, we obtain the $\nu_{n,0}$ and $\nu_{n,2}$ in Fig. \ref{fig:HD 5608 1}, and the corner diagram of asymptotic fit parameters in Fig. \ref{fig:HD 5608 3}.

  The $\nu _{max}$ is calculated to be 169.532 $\pm$ 0.614 $\mu$Hz and $\Delta \nu$ to be 12.348 $\pm$ 0.146 $\mu$Hz. Based on the $T_{eff}$ given by the paper \citep{stassun2017accurate} , which is 4929 $\pm$ 32 K, and on the basis of Eqs.\ref{equation:5} to \ref{equation:8} in Sec.\ref{sec:Seismology Analysis}, the physical parameters of the host star HD 5608 can be calculated. HD 5608 has the asteroseismic mass of 1.731 $\pm$ 0.087 $M_{\odot } $, the asteroseismic radius of 5.914 $\pm$ 0.144 $R_{\odot } $, the surface gravity of 3.134 $\pm$ 0.003 $\log_{10}{\mathrm{(cm/s^2)}} $, and the luminosity of 1.243 $\pm$ 0.024 $\log_{10}{L_{\odot }} $ calculated from these parameters. 

  From the parameters given in articles in Table \ref{tab:HD 5608}, the stellar parameters obtained using the asteroseismic analysis are in general agreement, and the stellar parameters directly determine the planetary parameters, so from Eqs. \ref{equation:10} and \ref{equation:11}, the changed planetary parameters can be obtained. Since the planet's transit signal was not available, the planet's orbital inclination could not be detected, and thus the planet's specific mass could not be confirmed, but the lower limit of the planet's mass, or minimum mass, $M_{p}\sin i$, could be calculated from the radial velocity data.The changed $M_{p}\sin i$ is 451.6 $\pm$ 28.9 $M_{\oplus }$ ($M_{\oplus }$ is the mass of the earth), which translates into the Jupiter mass of 1.421 $\pm$ 0.091 $M_{Jupiter}$. The orbital semi-major axis is 1.974 $\pm$ 0.035 AU (Astronomical Units).

  \begin{splitdeluxetable*}{lccccccccBcccccccc}
  \renewcommand{\arraystretch}{1.5}
  \tabletypesize{\scriptsize}
  \tablewidth{0pt} 
  \tablecaption{Parameters of system HD 5608 \label{tab:HD 5608}}
  \tablehead{
  \colhead{}  &
  \multicolumn{8}{c}{Parameters of Host Star HD 5608}  & \multicolumn{8}{c}{Parameters of Planet HD 5608 b}\\
  \cline{2-17}
  \colhead{Source} & \colhead{$T_{eff}$} & \colhead{$L_{\ast }$} & \colhead{$v\sin{i}$} & \colhead{$M_{\ast }$} & \colhead{$R_{\ast }$} & \colhead{[Fe/H]} & \colhead{$\log_{}{g} $} & \colhead{Spectral Type} &
  \colhead{$T_{p}$} & \colhead{$\omega$} & \colhead{$M_{p}\sin i$} & \colhead{$M_{p}\sin i$} &
  \colhead{e} & \colhead{P} & \colhead{K} & \colhead{a}\\
  \colhead{} & \colhead{K} & \colhead{$\log_{10}{L_{\odot }} $} & \colhead{km/s} & \colhead{$M_{\odot } $} & \colhead{$R_{\odot } $}  & \colhead{dex}  & \colhead{$\log_{10}{\mathrm{(cm/s^2)}} $} & \colhead{} & \colhead{days} & \colhead{degree} & \colhead{$M_{\oplus }$} & \colhead{$M_{Jupiter}$} & \colhead{} & \colhead{days} & \colhead{m/s} & \colhead{AU}
  } 
  \startdata 
  $\text{\citep{teng2023revisiting}}$&4807& $1.155_{-0.113}^{+0.109}$ &  \nodata &    1.29$\pm      0.23$& $5.43_{-0.62}^{+0.70}$ &$0.04_{-0.10}^{+0.07}$&     3.06$\pm      0.08$&    K0 IV&  $2453936.4_{-721.7}^{+27.8}$&     $-91.90_{-69.10}^{+15.40}$&    $371.2_{-18.1}^{+19.7}$&     $1.168_{-0.057}^{+0.062}$&  $0.110_{-0.080}^{+0.029}$&  $768.70_{-1.67}^{+4.72}$  &   $21.95_{-1.14}^{+1.17}$ &  $1.790_{-0.003}^{+0.007}$\\ 
  $\text{\citep{luhn2019retired}}$& 4877& \nodata & \nodata&  1.53 & 5.14 &\nodata&  3.19 &  \nodata &      2452420$\pm      110$&     294$\pm      54$& 534.3$\pm    25.7$&     1.681$\pm      0.081$& 0.056$\pm      0.039$ & 779.9 $\pm  4.9$ & 28.0$\pm1.4$ & 1.911 $\pm 0.088$\\
  $\text{\citep{stassun2017accurate}}$& 4929 $\pm 32$& \nodata &   \nodata&     1.98$\pm 0.26$& 5.09 $\pm 0.16$  & 0.14 &    3.32$\pm 0.05$    &   \nodata & \nodata &     \nodata  &     524$\pm      57$&     1.65$\pm 0.18$& 0.19$\pm       0.06$ & 792.6$\pm 7.7$ & 23.5$\pm 1.6$ & 1.99$\pm 0.04$\\
  $\text{\citep{sato2012substellar}}$& 4854$\pm 25$& 1.179 &    1.37& 1.55$\pm     0.23$&  5.5$\pm 0.4$ & \nodata &   \nodata &    \nodata &    2452327$\pm      61$&    269$\pm      22$&   445 & 1.4 & 0.190$\pm     0.061$ & 792.6$\pm 7.7$ & 23.5$\pm 1.6$ & 1.9\\
  \hline
  {This work}& \nodata & 1.243$\pm 0.024$ &    \nodata  &  1.731$\pm      0.087$&  5.914 $\pm 0.144$& \nodata &    3.134$\pm     0.003$&     \nodata&      \nodata&    \nodata &     451.6$\pm      28.9$& 1.421$\pm      0.091$&\nodata &\nodata &\nodata & 1.974 $\pm 0.035$\\
  \enddata
  \tablecomments{Since this work does not involve new radial velocity data, the data in this work are consistent with the latest published work, except for the new results obtained from asteroseismology. Parameter tables for subsequent research subjects are similar. }
  \end{splitdeluxetable*}

  \subsection{7 CMa}\label{7 CMa}
  7 CMa (HD 47205, TIC 48237215) is a K1 III giant with 2 exoplanets in its system. 7 CMa b, first discovered in 2011 \citep{wittenmyer2011pan}, is a massive planet with a minimum mass of at least twice the mass of Jupiter, but it is much closer to its host star than Jupiter. In 2016, as more RV data were obtained, Wittenmyer et al. made further corrections to the planet's parameters \citep{wittenmyer2016pan}. In 2019, after adding more RV data, Luque et al. found objects that cause changes in radial velocities, not only 7 CMa b, but also a new object with a minimum mass close to the mass of Jupiter, 7 CMa c. The result from the stability analysis indicates that the two-planets are trapped in a low-eccentricity 4:3 mean-motion resonance \citep{luque2019precise}.

  7 CMa was observed by TESS in December 2018 and December 2022 as Sector 6 and Sector 33, respectively. After a similar process as in Sec.\ref{HD 5608}, we obtained the asteroseismic parameters of the star. The calculation gives the $\nu _{max}$ equal to 182.926 $\pm$ 1.559 $\mu$Hz and $\Delta \nu$ equal to 13.662 $\pm$ 0.033 $\mu$Hz. With the paper \citep{luque2019precise} giving $T_{eff}$ equal to $4826_{-55}^{+45}$ K, the asteroseismic mass of 7 CMa is calculated to be 1.439 $\pm$ 0.047 $M_{\odot } $, the asteroseismic radius to be 5.198 $\pm$ 0.060 $R_{\odot } $, the surface gravity to be 3.165 $\pm$ 0.005 $\log_{10}{\mathrm{(cm/s^2)}} $, and the luminosity to be 1.121 $\pm$ 0.022 $\log_{10}{L_{\odot }} $. 

  As the mass of the host star changes, so do the parameters of the two planets. Closer to the host star, 7 CMa b has a minimum mass of 1.940 $\pm$ 0.064 $M_{Jupiter}$ (or 616.5 $\pm$ 20.2 $M_{\oplus }$) and an orbital semi-major axis of 1.800 $\pm$ 0.033 AU, and the more distant planet, 7 CMa c, has a minimum mass of 0.912 $\pm$ 0.067 $M_{Jupiter}$ (or 289.9 $\pm$ 21.2 $M_{\oplus }$) and an orbital semi-major axis of 2.205 $\pm$ 0.086 AU. Since the work in this paper involves corrections to the planetary parameters due to variations in the mass of the host star, there is no impact on the conclusions of the orbital resonance of the two planets in this system given the available radial velocity observations.

  \begin{figure}[ht!]

  \includegraphics[width=0.6\linewidth]{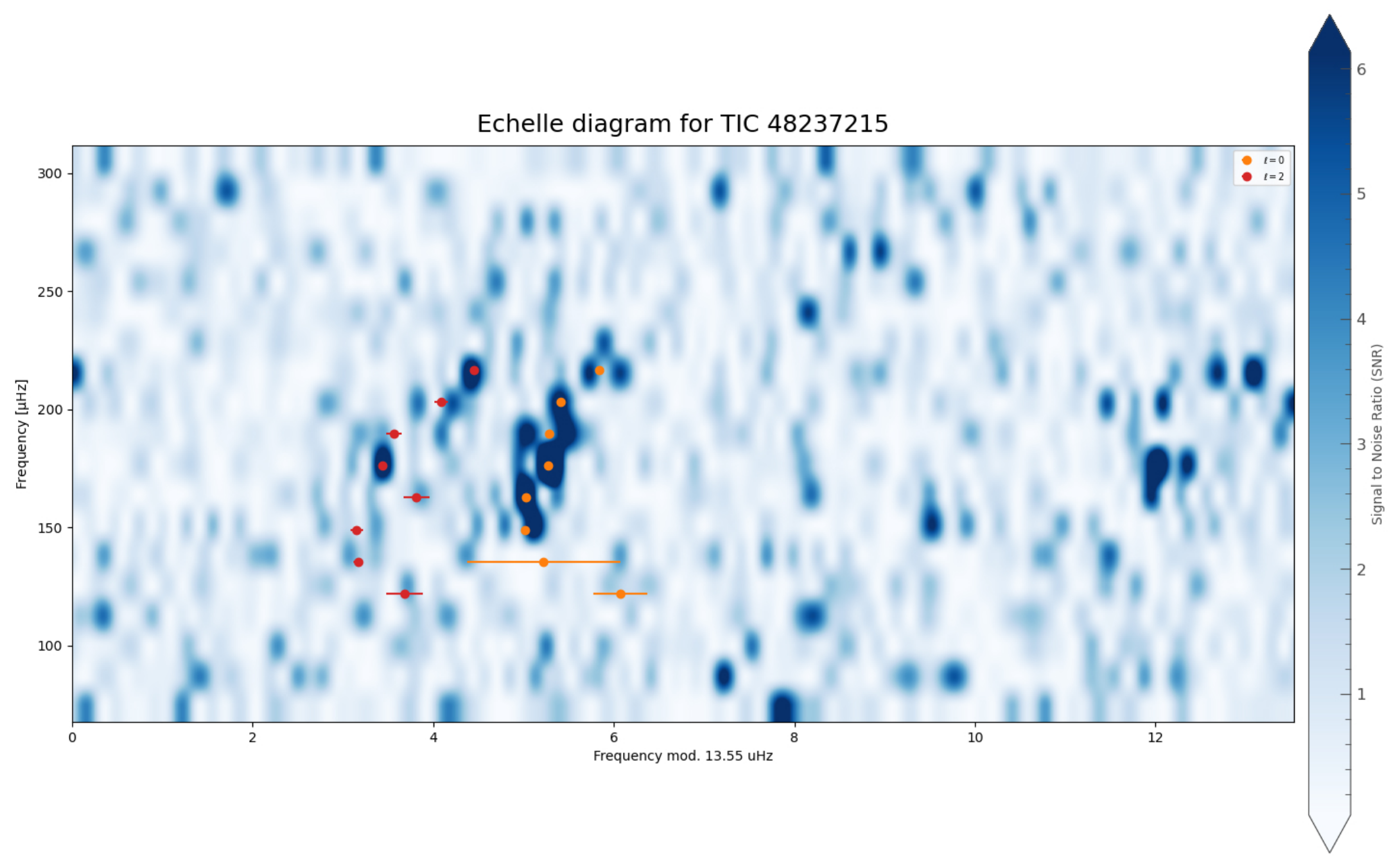}
  \centering
  \caption{7 CMa Echelle diagram of smoothed SNR spectral estimate. Orange dots with error bars mark where the radial $l=0$ mode is located, and red dots with error bars mark where the quadrupole $l=2$ mode is located. The $\Delta \nu$ value here is not an actual observed value computed.
  \label{fig:7 CMa 1}}

  \centering
  \includegraphics[width=0.6\linewidth]{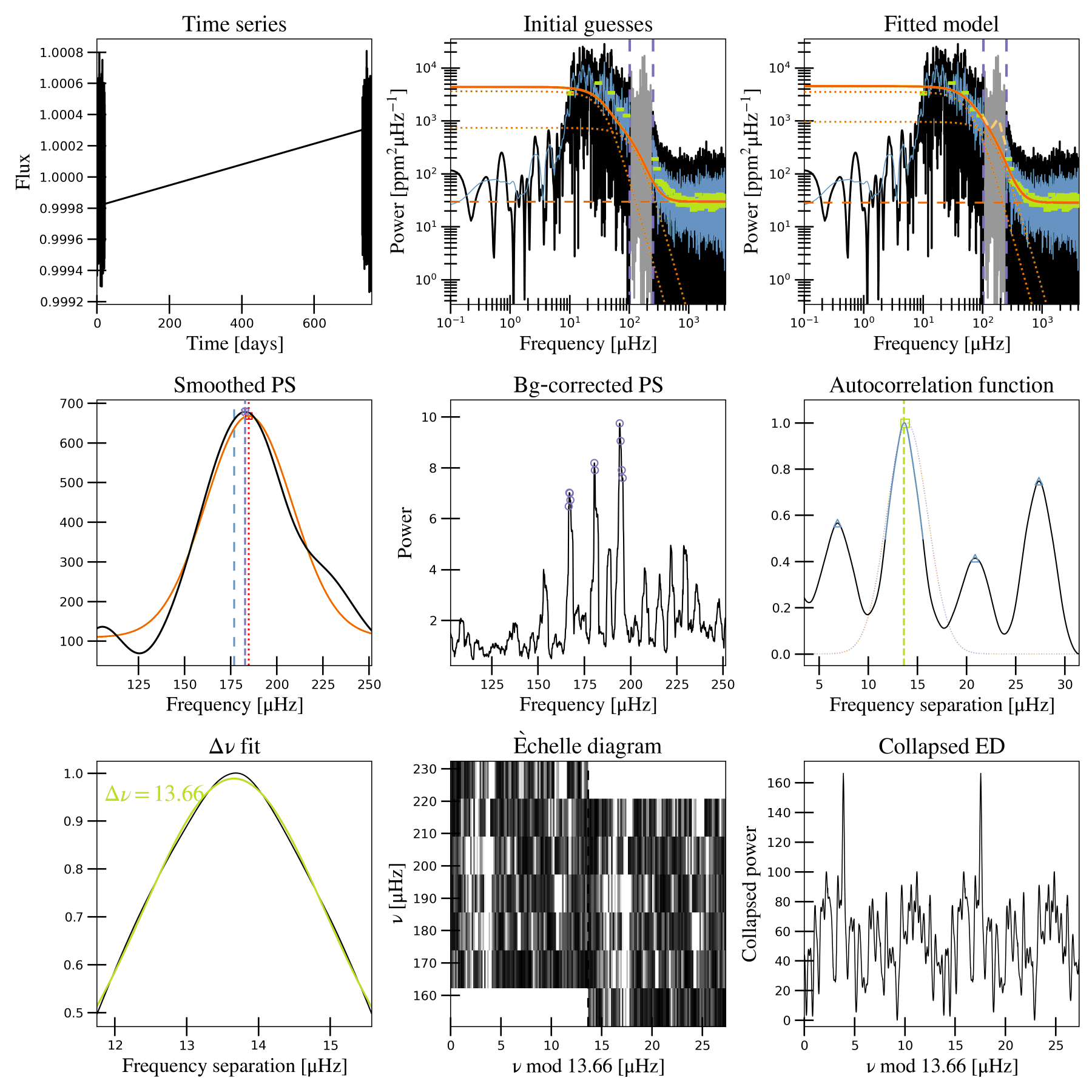}
  \caption{7 CMa global asteroseismic parameter fitting.
  \label{fig:7 CMa 2}}

  \end{figure}

  \begin{splitdeluxetable*}{lccccccccBccccccccBccccccc}
  \renewcommand{\arraystretch}{1.5}
  \tabletypesize{\scriptsize}
  \tablewidth{0pt} 
  \tablecaption{Parameters of system 7 CMa \label{tab:7 CMa}}
  \tablehead{
  \colhead{}  &
  \multicolumn{8}{c}{Parameters of Host Star 7 CMa}  & \multicolumn{8}{c}{Parameters of Planet 7 CMa b} & \multicolumn{7}{c}{Parameters of Planet 7 CMa c}\\
  \cline{2-24}
  \colhead{Source} & \colhead{$T_{eff}$} & \colhead{$L_{\ast }$} & \colhead{$v\sin{i}$} & \colhead{$M_{\ast }$} & \colhead{$R_{\ast }$} & \colhead{[Fe/H]} & \colhead{$\log_{}{g} $} & \colhead{Spectral Type} &
  \colhead{$T_{p}$} & \colhead{$\omega$} & \colhead{$M_{p}\sin i$} & \colhead{$M_{p}\sin i$} &
  \colhead{e} & \colhead{P} & \colhead{K} & \colhead{a} & \colhead{$\omega$} & \colhead{$M_{p}\sin i$} & \colhead{$M_{p}\sin i$} &
  \colhead{e} & \colhead{P} & \colhead{K} & \colhead{a}\\
  \colhead{} & \colhead{K} & \colhead{$\log_{10}{L_{\odot }} $} & \colhead{km/s} & \colhead{$M_{\odot } $} & \colhead{$R_{\odot } $}  & \colhead{dex}  & \colhead{$\log_{10}{\mathrm{(cm/s^2)}} $} & \colhead{} & \colhead{days} & \colhead{degree} & \colhead{$M_{\oplus }$} & \colhead{$M_{Jupiter}$} & \colhead{} & \colhead{days} & \colhead{m/s} & \colhead{AU} & \colhead{degree} & \colhead{$M_{\oplus }$} & \colhead{$M_{Jupiter}$} & \colhead{} & \colhead{days} & \colhead{m/s} & \colhead{AU}
  } 
  \startdata 
  $\text{\citep{luque2019precise}}$&$4826_{-55}^{+45}$& $1.063_{-0.008}^{+0.012}$ &  \nodata &   $1.34_{-0.12}^{+0.11}   $& $4.87_{-0.14}^{+0.17}$ & 0.21$\pm 0.10$& $3.19_{-0.07}^{+0.06}$&  \nodata & \nodata &     $165.3_{-70.8}^{+5.1}$ &    $588_{-13}^{+19}$&     $1.85_{-0.04}^{+0.06}$&  0.06$\pm  0.03$&  $735.1_{-1.0}^{+14.8}$  &   $34.3_{-0.9}^{+1.2}$ &  $1.758_{-0.001}^{+0.024}$  & $233.5_{-40.2}^{+7.7}$ & 277$\pm 19$ & 0.87$\pm 0.06$ & $0.08_{-0.04}^{+0.05}$ & $996.0_{-52.4}^{+1.5}$ & $14.9_{-1.1}^{+0.9}$ & $2.153_{-0.080}^{+0.003}$\\ 
  $\text{\citep{wittenmyer2016pan}}$& \nodata & \nodata & \nodata & \nodata & \nodata & \nodata & \nodata &  \nodata &      2454093$\pm      34$&     77$\pm      14$& 782$\pm    44$&     2.46$\pm      0.14$& 0.22$\pm      0.07$ & 796.0$\pm  7.4$ & 41.8$\pm2.4$ & 1.93 $\pm 0.01$& \nodata & \nodata & \nodata & \nodata & \nodata & \nodata & \nodata \\
  $\text{\citep{wittenmyer2011pan}}$& 4792 $\pm 100$& $1.053_{-0.012}^{+0.011}$ &   1.15 &     1.52$\pm 0.30$& 2.3 $\pm 0.1$  & 0.21$\pm 0.10$ &    3.25$\pm 0.10$    & K1 III & 2455520$\pm 89$ &  12$\pm 41$  &     826.3$\pm      190.7$&     2.6$\pm 0.6$& 0.14$\pm       0.06$ & 763$\pm 17$ & 44.9$\pm 4.0$ & 1.9$\pm 0.1$& \nodata & \nodata & \nodata & \nodata & \nodata & \nodata & \nodata \\
  \hline
  {This work}& \nodata & 1.121$\pm 0.022$ &    \nodata  &  1.439$\pm      0.047$&  5.198$\pm 0.060$& \nodata &    3.165$\pm     0.005$&     \nodata&      \nodata&    \nodata &     616.5$\pm      20.2$& 1.940$\pm      0.064$&\nodata &\nodata &\nodata & 1.800 $\pm 0.033$&\nodata & 289.9$\pm21.2$ & 0.912$\pm0.067$ &\nodata&\nodata&\nodata & 2.205$\pm0.086$\\
  \enddata
  \end{splitdeluxetable*}

  \subsection{HD 33844}\label{HD 33844}

  HD 33844 (TIC 169534187) is a K0 III giant with 2 exoplanets in its system. Both planets were discovered by Wittenmyer et al. in 2016 via the radial velocity method \citep{wittenmyer2016pan2}. Detailed N-body dynamical simulations show that the two planets have remained on stable orbits for more than $10^{6}$ years for low eccentricities and are most likely trapped in a mutual 3:5 mean motion resonance \citep{wittenmyer2016pan2}. And then in 2017, Stassun et al. redetermined the radii, masses, and other parameters of nearly 500 host stars through Gaia's survey data, which in turn corrected the relevant parameters of the planet \citep{stassun2017accurate}, including HD 33844.

  In this work, the asteroseismic parameters of the host star were obtained by analyzing TESS observations of HD 33844 in November 2018 (Sector 5) and December 2020 (Sector 32). The $\nu _{max}$ is equal to 165.951 $\pm$ 1.171 $\mu$Hz, the $\Delta \nu$ is equal to 12.665 $\pm$ 0.120 $\mu$Hz. The paper gives HD 33844 an $T_{eff}$ of 4861 $\pm$ 100 K \citep{stassun2017accurate}, which in turn calculates the star's asteroseismic mass to be 1.471 $\pm$ 0.080 $M_{\odot } $, its asteroseismic radius to be 5.507 $\pm$ 0.127 $R_{\odot } $, its surface gravity to be 3.135 $\pm$ 0.006 $\log_{10}{\mathrm{(cm/s^2)}} $, and from the temperature and radius it can be further calculated to have a luminosity of 1.183 $\pm$ 0.048 $\log_{10}{L_{\odot }} $.

  The minimum masses and orbital semi-major axis of the two planets are reduced because the stellar mass obtained by asteroseismology is smaller than that given in papers. HD 33844 b, which is closer to the host star, has a minimum mass of 1.726 $\pm$ 0.145 $M_{Jupiter}$ (548.4 $\pm$ 46.2 $M_{\oplus }$) and an orbital semi-major axis of 1.501 $\pm$ 0.047 AU, and HD 33844 c, which is farther from the host star, has a minimum mass of 1.541 $\pm$ 0.182 $M_{Jupiter}$ (489.7 $\pm$ 57.8 $M_{\oplus }$) and an orbital semi-major axis of 2.111 $\pm$ 0.077 AU.

  \begin{figure}[ht!]

  \includegraphics[width=0.6\linewidth]{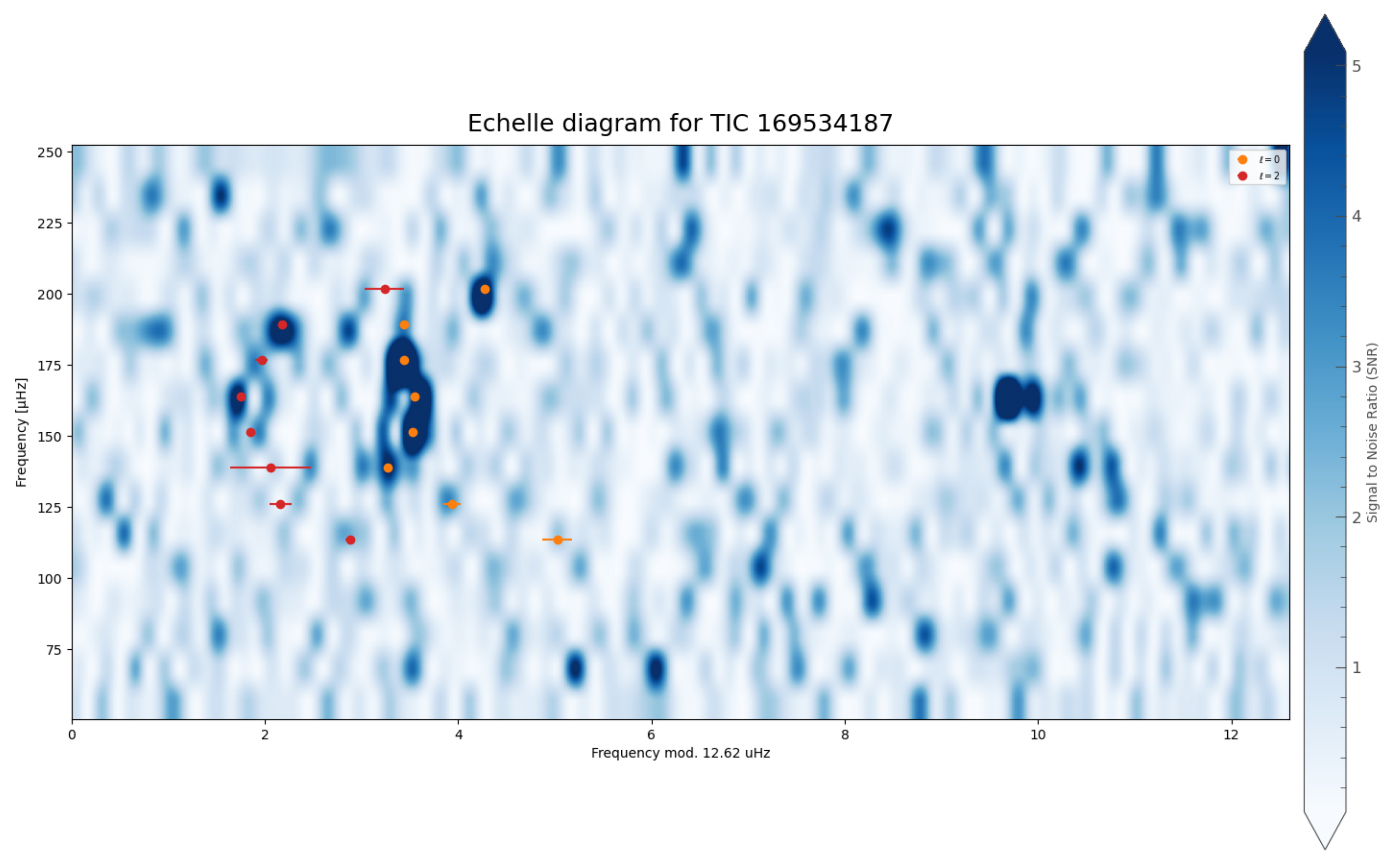}
  \centering
  \caption{HD 33844 Echelle diagram of smoothed SNR spectral estimate. Orange dots with error bars mark where the radial $l=0$ mode is located, and red dots with error bars mark where the quadrupole $l=2$ mode is located. The $\Delta \nu$ value here is not an actual observed value computed.
  \label{fig:HD 33844 1}}

  \centering
  \includegraphics[width=0.6\linewidth]{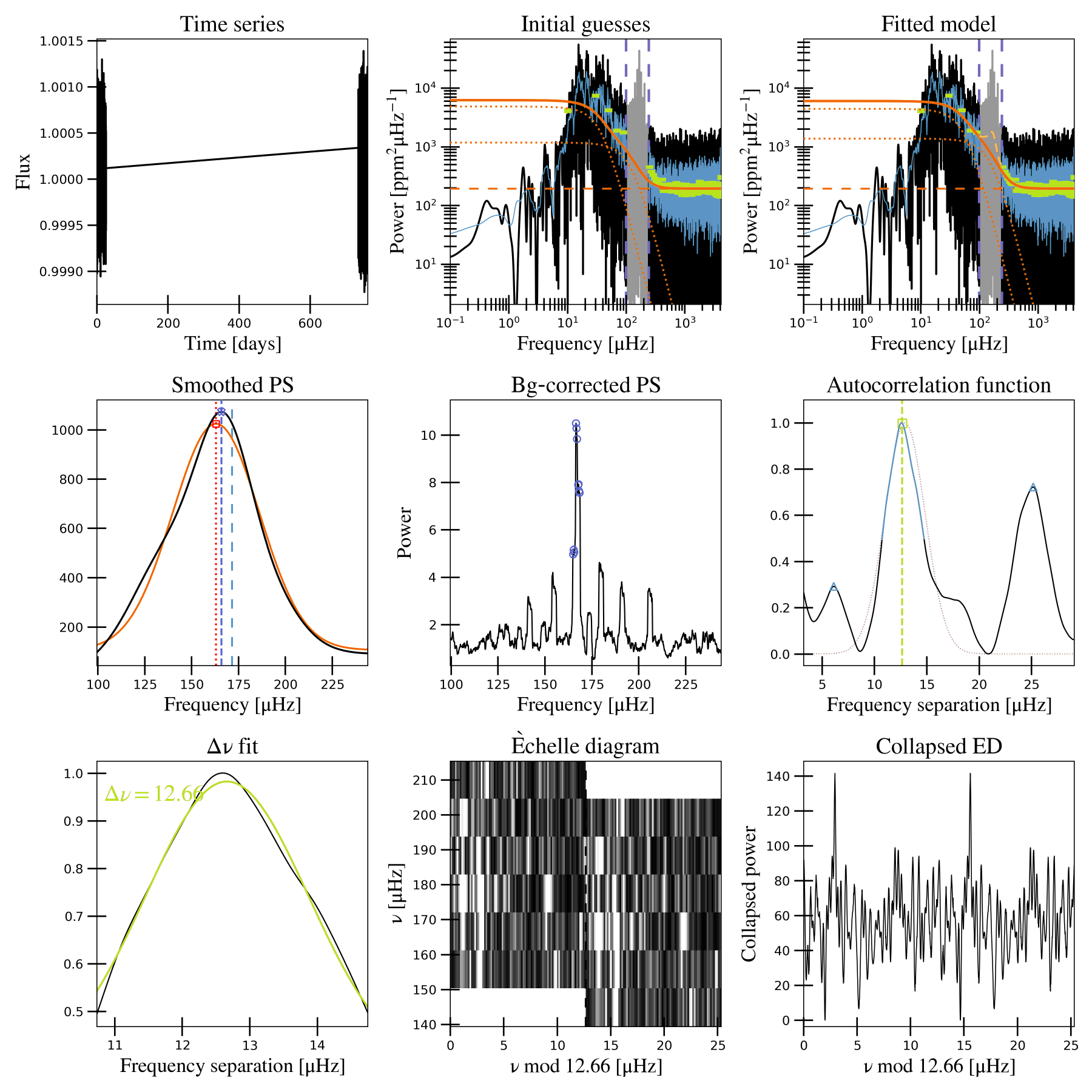}
  \caption{HD 33844 global asteroseismic parameter fitting.
  \label{fig:HD 33844 2}}

  \end{figure}

  \begin{splitdeluxetable*}{lccccccccBccccccccBcccccccc}
  \renewcommand{\arraystretch}{1.5}
  \tabletypesize{\scriptsize}
  \tablewidth{0pt} 
  \tablecaption{Parameters of system HD 33844 \label{tab:HD 33844}}
  \tablehead{
  \colhead{}  &
  \multicolumn{8}{c}{Parameters of Host Star HD 33844}  & \multicolumn{8}{c}{Parameters of Planet HD 33844 b} & \multicolumn{8}{c}{Parameters of Planet HD 33844 c}\\
  \cline{2-25}
  \colhead{Source} & \colhead{$T_{eff}$} & \colhead{$L_{\ast }$} & \colhead{$v\sin{i}$} & \colhead{$M_{\ast }$} & \colhead{$R_{\ast }$} & \colhead{[Fe/H]} & \colhead{$\log_{}{g} $} & \colhead{Spectral Type} &
  \colhead{$T_{p}$} & \colhead{$\omega$} & \colhead{$M_{p}\sin i$} & \colhead{$M_{p}\sin i$} &
  \colhead{e} & \colhead{P} & \colhead{K} & \colhead{a} &\colhead{$T_{p}$}& \colhead{$\omega$} & \colhead{$M_{p}\sin i$} & \colhead{$M_{p}\sin i$} &
  \colhead{e} & \colhead{P} & \colhead{K} & \colhead{a}\\
  \colhead{} & \colhead{K} & \colhead{$\log_{10}{L_{\odot }} $} & \colhead{km/s} & \colhead{$M_{\odot } $} & \colhead{$R_{\odot } $}  & \colhead{dex}  & \colhead{$\log_{10}{\mathrm{(cm/s^2)}} $} & \colhead{} & \colhead{days} & \colhead{degree} & \colhead{$M_{\oplus }$} & \colhead{$M_{Jupiter}$} & \colhead{} & \colhead{days} & \colhead{m/s} & \colhead{AU} &\colhead{days}& \colhead{degree} & \colhead{$M_{\oplus }$} & \colhead{$M_{Jupiter}$} & \colhead{} & \colhead{days} & \colhead{m/s} & \colhead{AU}
  } 
  \startdata 
  $\text{\citep{stassun2017accurate}}$&4861$\pm100$& \nodata &  \nodata &   1.84$\pm0.39$& 5.39$\pm0.28$ & 0.27& 3.24$\pm0.08$&  \nodata & \nodata &     \nodata &    639$\pm99$&     2.01$\pm0.31$&  0.15$\pm  0.07$&  551.4$\pm7.8$  &   33.5$\pm2.0$ &  1.60$\pm0.06$  & \nodata & \nodata & \nodata & \nodata & \nodata& \nodata & \nodata& \nodata\\ 
  $\text{\citep{wittenmyer2016pan2}}$& 4861$\pm100$ & $1.15_{-0.06}^{+0.05}$ & $<1$ & 1.78$\pm0.18$ & 5.29$\pm0.41$ & 0.27$\pm0.09$ & 3.24$\pm0.08$ &  K0 III &      2454609$\pm      41$&     211$\pm      28$& 623$\pm    38$&     1.96$\pm      0.12$& 0.15$\pm      0.07$ & 551.4$\pm  7.8$ & 33.5$\pm2.0$ & 1.60 $\pm 0.02$& 2454544$\pm164$ & 71.0$\pm67.0$ & 556$\pm57$ & 1.75$\pm0.18$ & 0.13$\pm0.10$ & 916.0$\pm29.5$ & 25.4$\pm2.9$ & 2.24$\pm0.05$ \\
  \hline
  {This work}& \nodata & 1.183$\pm 0.048$ &    \nodata  &  1.471$\pm      0.080$&  5.507$\pm 0.127$& \nodata &    3.125$\pm     0.006$&     \nodata&      \nodata&    \nodata &     548.4$\pm      46.2$& 1.726$\pm      0.145$&\nodata &\nodata &\nodata & 1.501 $\pm 0.047$& \nodata&\nodata & 489.7$\pm57.8$ & 1.541$\pm0.182$ &\nodata&\nodata&\nodata & 2.111$\pm0.077$\\
  \enddata
  \end{splitdeluxetable*}

  \subsection{HIP 67851}\label{HIP 67851}
  The two planets of K0 III giant HIP 67851 were independently discovered in 2015 by Jones et al. \citep{jones2015giant} and Wittenmyer et al \citep{wittenmyer2015pan}. giving similar results for planetary parameters. In the same year, the planetary parameters were corrected based on richer RV data by Jones et al \citep{jones2015planetary}. And in 2022, Feng et al. \citep{feng20223d} made a fit to the planet's orbit based on parallax data from the Gaia satellite, combined with existing RV data, to find the orbital inclination, which led to the determination of the mass of HIP 67851 c. However, it actually gave a very large error. In the present work, the mass correction for HIP 67851 c is given in Table \ref{tab:HIP 67851} as a minimum mass only, due to the fact that changing the mass of the host star, which directly affects the orbital semi-major axis, does not allow for a direct application of the combined astrometry to correct the planetary mass.

  HIP 67851 was observed by TESS in May 2019 (Sector 11), May 2021 (Sector 38), and April 2023 (Sector 64) and was analyzed by $pySYD$ to obtain $\nu _{max}$ equal to 134.433 $\pm$ 0.638 $\mu$Hz and $\Delta \nu$ equal to 10.985 $\pm$ 0.027 $\mu$Hz. The paper \citep{jones2015giant} gives $T_{eff}$ equal to 4890 $\pm$ 100 K, and substituting the formulae yields the star's asteroseismic mass of 1.394  $\pm$ 0.052 $M_{\odot } $ and a asteroseismic radius of 5.948 $\pm$ 0.077 $R_{\odot } $, and a surface gravity of 3.034 $\pm$ 0.005 $\log_{10}{\mathrm{(cm/s^2)}} $, from which the star's luminosity is calculated to be 1.261 $\pm$ 0.045 $\log_{10}{L_{\odot }} $.

  On the basis of the corrected stellar mass, HIP 67851 b has a minimum mass of 1.243 $\pm$ 0.139 $M_{Jupiter}$, which translates to 441.6 $\pm$ 56.8 $M_{\oplus }$, and it is at a distance of 0.437 $\pm$ 0.020 AU from its host star.Another planet farther away from its host star, HIP 67851 c, has a minimum mass of 5.387 $\pm$ 0.699 $M_{Jupiter}$, which translates to 1712.0 $\pm$ 222.0 $M_{\oplus }$, and its orbital semi-major axis is 3.626 $\pm$ 0.223 AU.

  \begin{figure}[ht!]

  \includegraphics[width=0.6\linewidth]{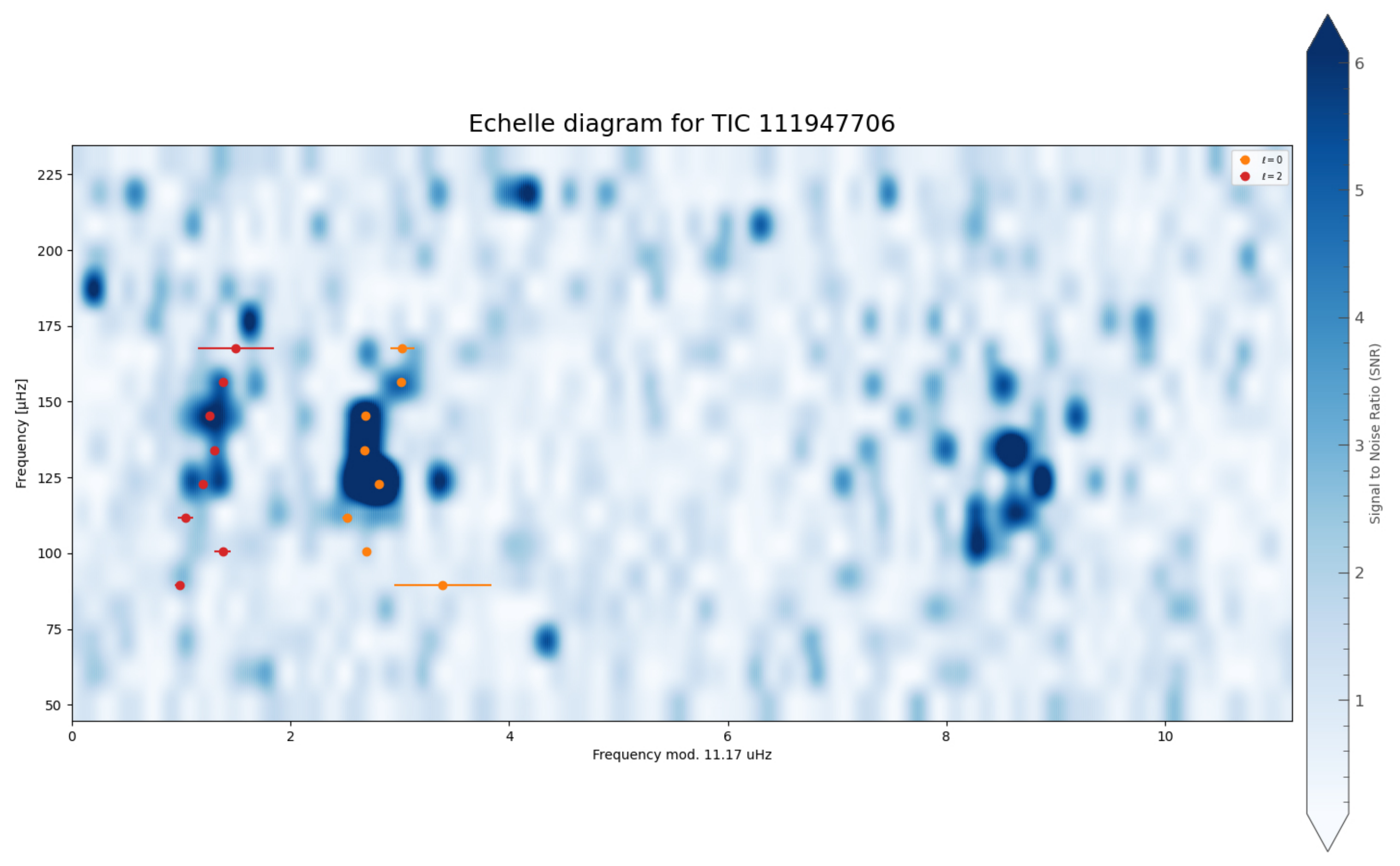}
  \centering
  \caption{HIP 67851 Echelle diagram of smoothed SNR spectral estimate. Orange dots with error bars mark where the radial $l=0$ mode is located, and red dots with error bars mark where the quadrupole $l=2$ mode is located.  The $\Delta \nu$ value here is not an actual observed value computed.
  \label{fig:HIP 67851 1}}

  \centering
  \includegraphics[width=0.6\linewidth]{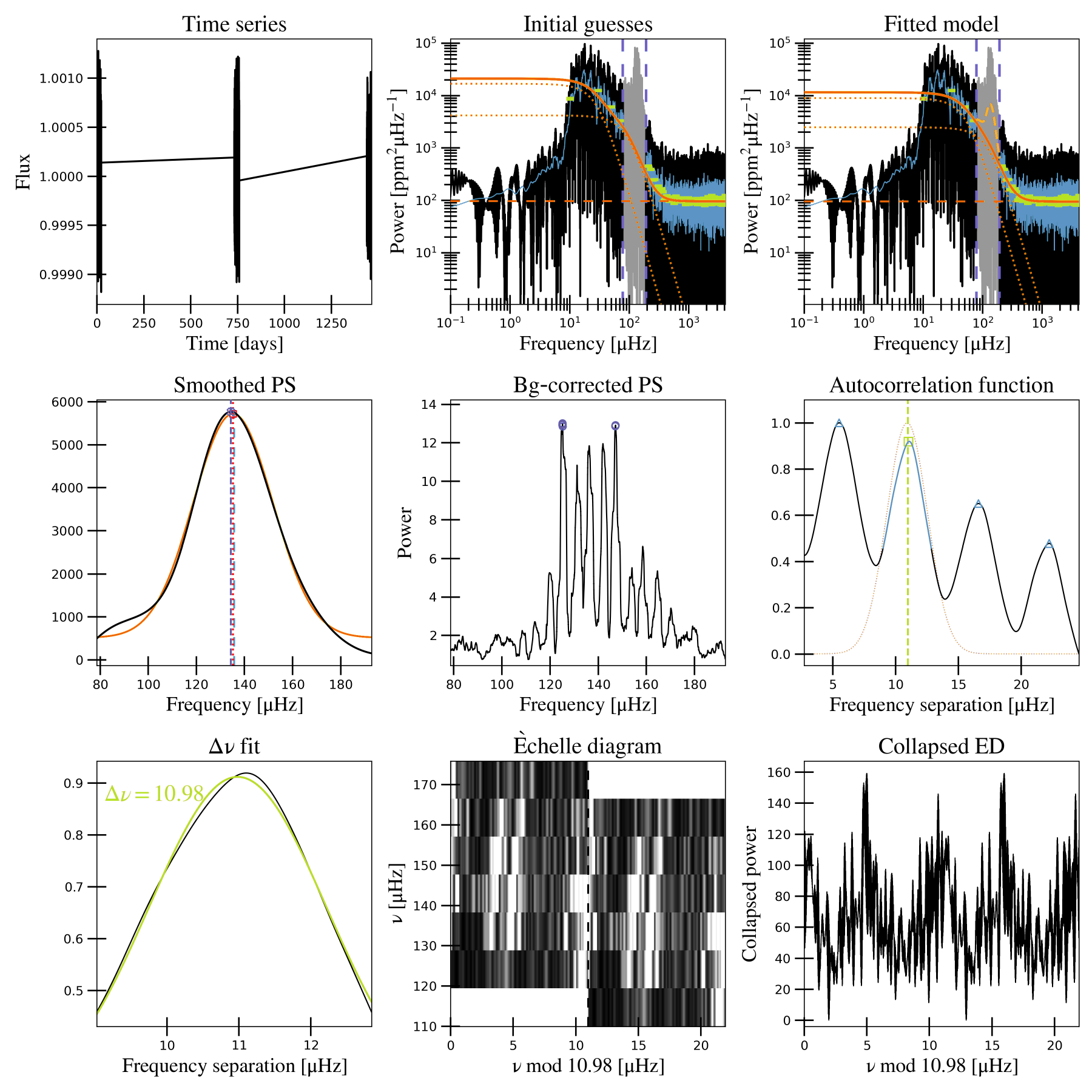}
  \caption{HIP 67851 global asteroseismic parameter fitting.
  \label{fig:HIP 67851 2}}

  \end{figure}

  \begin{splitdeluxetable*}{lccccccccBccccccccBccccccccc}
  \renewcommand{\arraystretch}{1.5}
  \tabletypesize{\scriptsize}
  \tablewidth{0pt} 
  \tablecaption{Parameters of system HIP 67851 \label{tab:HIP 67851}}
  \tablehead{
  \colhead{}  &
  \multicolumn{8}{c}{Parameters of Host Star HIP 67851}  & \multicolumn{8}{c}{Parameters of Planet HIP 67851 b} & \multicolumn{9}{c}{Parameters of Planet HIP 67851 c}\\
  \cline{2-26}
  \colhead{Source} & \colhead{$T_{eff}$} & \colhead{$L_{\ast }$} & \colhead{$v\sin{i}$} & \colhead{$M_{\ast }$} & \colhead{$R_{\ast }$} & \colhead{[Fe/H]} & \colhead{$\log_{}{g} $} & \colhead{Spectral Type} &
  \colhead{$T_{p}$} & \colhead{$\omega$} & \colhead{$M_{p}\sin i$} & \colhead{$M_{p}\sin i$} &
  \colhead{e} & \colhead{P} & \colhead{K} & \colhead{a} &\colhead{$T_{p}$}& \colhead{$\omega$} & \colhead{$M_{p}\sin i$} & \colhead{$M_{p}\sin i$} &
  \colhead{e} & \colhead{P} & \colhead{K} & \colhead{a} &\colhead{i}\\
  \colhead{} & \colhead{K} & \colhead{$\log_{10}{L_{\odot }} $} & \colhead{km/s} & \colhead{$M_{\odot } $} & \colhead{$R_{\odot } $}  & \colhead{dex}  & \colhead{$\log_{10}{\mathrm{(cm/s^2)}} $} & \colhead{} & \colhead{days} & \colhead{degree} & \colhead{$M_{\oplus }$} & \colhead{$M_{Jupiter}$} & \colhead{} & \colhead{days} & \colhead{m/s} & \colhead{AU} &\colhead{days}& \colhead{degree} & \colhead{$M_{\oplus }$} & \colhead{$M_{Jupiter}$} & \colhead{} & \colhead{days} & \colhead{m/s} & \colhead{AU} & \colhead{degree}
  } 
  \startdata 
  $\text{\citep{feng20223d}}$&\nodata& \nodata &  \nodata &   1.896$\pm0.109$& \nodata& \nodata& \nodata&  \nodata & $2452989.1_{-1.8}^{+78.7}$ &     $101.9_{-25.4}^{+35.6}$ &    $489.4_{-22.9}^{+22.6}$& $1.568_{-0.072}^{+0.071}$&  $0.063_{-0.032}^{+0.031}$&  $88.876_{-0.11971}^{+0.11842}$  &  $46.664_{-1.122}^{+1.162}$ &  $0.482_{-0.010}^{+0.009}$ & $2451730_{-560}^{295}$ & $321.3_{-13.2}^{+11.6}$ &$\star$ $2205_{-165}^{+650}$  & $\star$ $6.937_{-0.518}^{+2.045}$ & $0.356_{-0.085}^{+0.112}$& $3923_{-330}^{+616}$ & $62.29_{-7.723}^{+7.574}$& $6.03_{-0.36}^{+0.65}$ & $89.9_{-24.7}^{+28.4}$\\ 
  $\text{\citep{jones2015giant}}$& 4890$\pm100$ & $1.244_{-0.071}^{+0.061}$ & 1.8$\pm0.9$ & 1.63$\pm0.22$ & 5.92$\pm0.44$ & 0.00$\pm0.10$ & 3.15$\pm0.20$ &  K0 III &      2452997.8$\pm      16.7$&     138.1$\pm      60.0$& 439$\pm    48$&     1.38$\pm      0.15$& 0.05$\pm      0.04$ & 88.9$\pm  0.1$ & 45.5$\pm1.6$ & 0.46 $\pm 0.02$& 2452684.1$\pm235.7$ & 166.5$\pm20.5$ & 1901$\pm242$ & 5.98$\pm0.76$ & 0.17$\pm0.06$ & 2131.8$\pm88.3$ & 69.0$\pm3.3$ & 3.82$\pm0.23$ &\nodata \\
  $\text{\citep{wittenmyer2015pan}}$ & 4805$\pm100$ & 1.207$\pm0.021$ &2&  
    1.30$\pm0.18$ & 5.87$\pm0.29$ &-0.03$\pm0.10$ & 3.04$\pm0.10$ & K0 III & 2455329.8$\pm2.6$ & 244$\pm10$ & 429.05$\pm54.03$ & 1.35$\pm0.17$ & 0.17$\pm0.04$ &89.06$\pm0.10$ & 52.5$\pm2.2$ & 0.426$\pm0.020$ & 2453290$\pm59$ & 136$\pm11$ & 1233.13$\pm174.80$ & 3.88$\pm0.55$ & 0.20$\pm0.05$ & 1626$\pm26$ & 57.4$\pm3.1$ & 2.96$\pm0.16$ & \nodata \\
  $\text{\citep{jones2015planetary}}$ & 4890$\pm100$ & $1.244_{-0.071}^{+0.061}$ & 1.8$\pm0.9$ & 1.63$\pm0.22$ & 5.92$\pm0.44$ & 0.00$\pm0.10$ & 3.15$\pm0.20$ &  K0 III & 2455296.6$\pm7.7$ & 87.7$\pm36.4$ & 457.66$\pm12.71$ & 1.44$\pm0.04$ & 0.09$\pm0.05$& 88.8$\pm0.2$ & 46.8$\pm2.0$ & 0.46$\pm0.01$ & \nodata&\nodata&$\dagger$ 2002.2&$\dagger$ 6.3&0&2200&\nodata&\nodata&\nodata\\
  \hline
  {This work}& \nodata & 1.261$\pm 0.045$ &    \nodata  &  1.394$\pm      0.052$&  5.948$\pm 0.077$& \nodata &    3.034$\pm     0.005$&     \nodata&      \nodata&    \nodata &     441.6$\pm      56.8$& 1.243$\pm      0.139$ &\nodata &\nodata &\nodata & 0.437 $\pm 0.020$& \nodata&\nodata & 1712.0$\pm222.0$ & 5.387$\pm0.699$ &\nodata&\nodata&\nodata & 3.626$\pm0.223$\\
  \enddata
  \tablecomments{ $\star$ Since the orbital inclination was obtained in the work of this article, the exact mass of the planet for this purpose is given here. }
  \tablecomments{ $\dagger$ This paper gives only one possible solution, so it is here only for reference.}
  \end{splitdeluxetable*}

  \section{Conclusion} \label{sec:conclusion}

  The asteroseismic parameters of the four stars analyzed in this work were obtained using the $pySYD$ package based on TESS data. These parameters were used to calculate the mass, radius, luminosity, and surface gravity of the host stars. This workflow has been validated and widely applied as a well-established method for measuring stellar parameters independent of spectroscopy. Observations by sky surveying satellites have provided a large amount of data on asteroseismology. However, due to the requirement of high precision and long observation periods for exoplanet discovery, there is a limited amount of asteroseismic data available for host stars of confirmed exoplanets.

  This work focuses on screening for late-evolved host stars. These stars have easily observable asteroseismic light variations, and the impact of expanding stars on the planets in the system is significant. After cross-referencing TESS observations with data on the host stars of confirmed exoplanets, we have identified dozens of late-evolved host stars with asteroseismic light variations that have not been analyzed, enumerating four of them in the present work and further correcting the exoplanet parameters in their systems. The results indicate that the parameters obtained by asteroseismology are close to those obtained by spectroscopy. Although we did not make any further observations of the radial velocities of these sources, we were able to obtain corrections to the exoplanet parameters based on previous work, which side-steps the importance of asteroseismology for exoplanet research work.

  Our present work is a preliminary phase. Future follow-up observations will reveal more about the system's evolution. The presence of exoplanets makes these late-evolved host stars particularly important. Asteroseismology provides a glimpse into the stars' interiors, making these planetary systems an invaluable natural laboratory for studying the future of our solar system.

  \section*{Acknowledgments}

  This work is supported by National Key R\&D Progrom of China (grant No:2022YFE0116800),the National Natural Science Foundation of China (No. 11933008 and No. 12103084), the basic research project of Yunnan Province (Grant No. 202301AT070352). This paper includes data collected by the TESS mission, which are publicly available from the Mikulski Archive for Space Telescopes (MAST)\footnote{https://mast.stsci.edu/}. Funding for the TESS mission is provided by the NASA Science Mission Directorate.

  \bibliography{sample631}{}
  \bibliographystyle{aasjournal}

  \section{Appendix}\label{sec:Appendix}

  \renewcommand\thetable{A1}

  \begin{table}[h!]
    \centering
    \caption{\textbf{$\nu_{n,0}$ and $\nu_{n,2}$ of 4 host stars}}
    \label{tab:freqs}
    \tabletypesize{\scriptsize}
    \tablewidth{0pt}
    \renewcommand{\arraystretch}{1.5}
    \begin{tabular}{cccccc}
      \hline
      \hline
      & & \textbf{HD 5608} & \textbf{7 CMa} & \textbf{HD 33844} & \textbf{HIP 67851} \\
      ID & $l$ & Frequency [$\mu Hz$] & Frequency [$\mu Hz$] & Frequency [$\mu Hz$] & Frequency [$\mu Hz$] \\
      \hline
      $f_{1}$  & 0 & 106.615$\pm0.370$ & 128.029$\pm0.299$ & 118.606$\pm0.156$ & 92.775$\pm0.447$\\
      $f_{2}$  & 0 & 118.447$\pm0.304$ & 140.734$\pm0.852$ & 130.134$\pm0.088$ & 103.245$\pm0.029$ \\
      $f_{3}$  & 0 & 131.716$\pm0.051$ & 154.079$\pm0.036$ & 142.093$\pm0.042$ & 114.243$\pm0.046$ \\
      $f_{4}$  & 0 & 144.290$\pm0.016$ & 167.643$\pm0.024$ & 154.969$\pm0.026$ & 125.708$\pm0.019$ \\
      $f_{5}$  & 0 & 157.041$\pm0.016$ & 181.440$\pm0.019$ & 167.609$\pm0.028$ & 136.753$\pm0.015$ \\
      $f_{6}$  & 0 & 169.843$\pm0.023$ & 195.003$\pm0.035$ & 180.115$\pm0.017$ & 147.930$\pm0.015$ \\
      $f_{7}$  & 0 & 182.545$\pm0.026$ & 208.674$\pm0.018$ & 192.736$\pm0.029$ & 159.433$\pm0.015$ \\
      $f_{8}$  & 0 & 195.515$\pm0.021$ & 222.661$\pm0.037$ & 206.190$\pm0.027$ & 170.612$\pm0.111$ \\
      $f_{9}$  & 2 & 105.122$\pm0.379$ & 125.646$\pm0.202$ & 116.459$\pm0.051$ & 90.367$\pm0.041$ \\
      $f_{10}$ & 2 & 117.633$\pm0.178$ & 138.681$\pm0.051$ & 128.361$\pm0.116$ & 101.933$\pm0.075$ \\
      $f_{11}$ & 2 & 130.185$\pm0.022$ & 152.216$\pm0.069$ & 140.875$\pm0.418$ & 112.767$\pm0.074$ \\
      $f_{12}$ & 2 & 143.231$\pm0.019$ & 166.428$\pm0.143$ & 153.288$\pm0.039$ & 124.097$\pm0.024$ \\
      $f_{13}$ & 2 & 155.453$\pm0.013$ & 179.602$\pm0.021$ & 165.813$\pm0.034$ & 135.376$\pm0.018$ \\
      $f_{14}$ & 2 & 168.369$\pm0.017$ & 193.279$\pm0.083$ & 178.642$\pm0.063$ & 146.504$\pm0.021$ \\
      $f_{15}$ & 2 & 181.437$\pm0.066$ & 207.358$\pm0.074$ & 191.473$\pm0.035$ & 157.800$\pm0.027$ \\
      $f_{16}$ & 2 & 194.386$\pm0.021$ & 221.264$\pm0.023$ & 205.155$\pm0.201$ & 169.089$\pm0.347$ \\
      \hline
    \end{tabular}
  \end{table}

  \section{Asymptotic Fit Corner Diagrams}

  \begin{figure}[ht!]
  \includegraphics[width=0.58\linewidth]{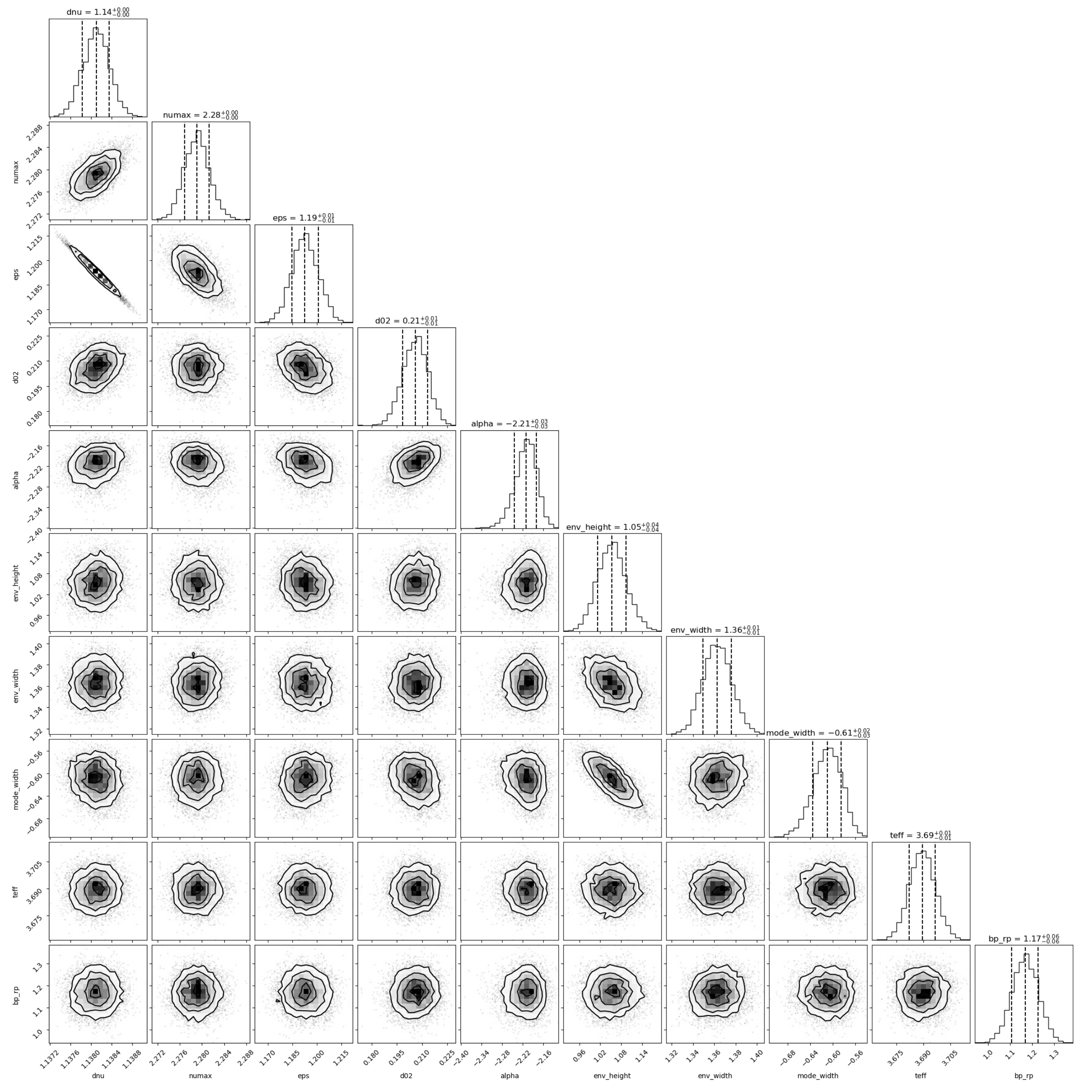}
  \centering
  \caption{7 CMa corner diagram of asymptotic fit parameters.
  \label{fig:7 CMa 3}}
  \end{figure}

  \begin{figure}[ht!]
  \includegraphics[width=0.58\linewidth]{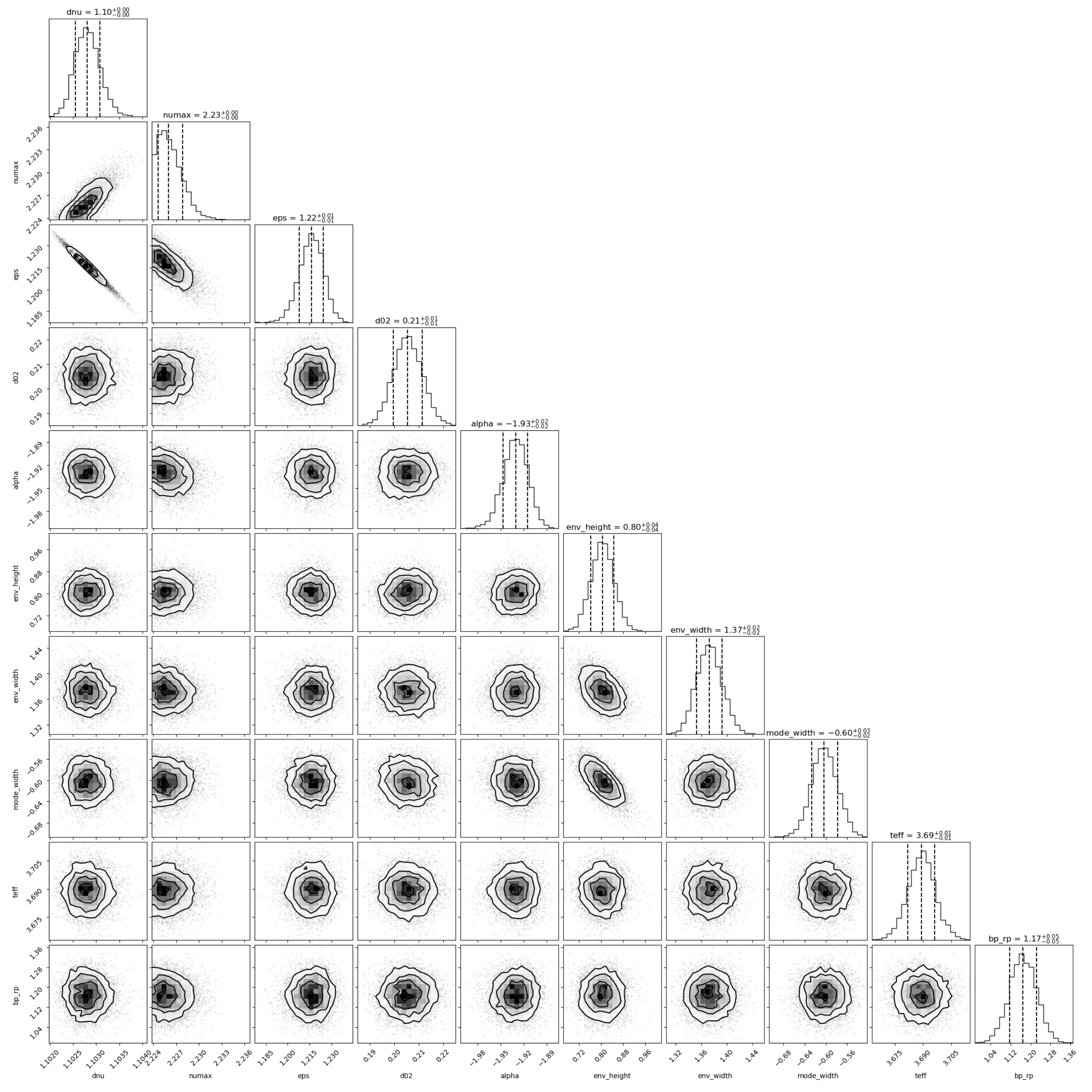}
  \centering
  \caption{HD 33844 corner diagram of asymptotic fit parameters.
  \label{fig:HD 33844 3}}
  \end{figure}

  \begin{figure}[ht!]
  \includegraphics[width=0.58\linewidth]{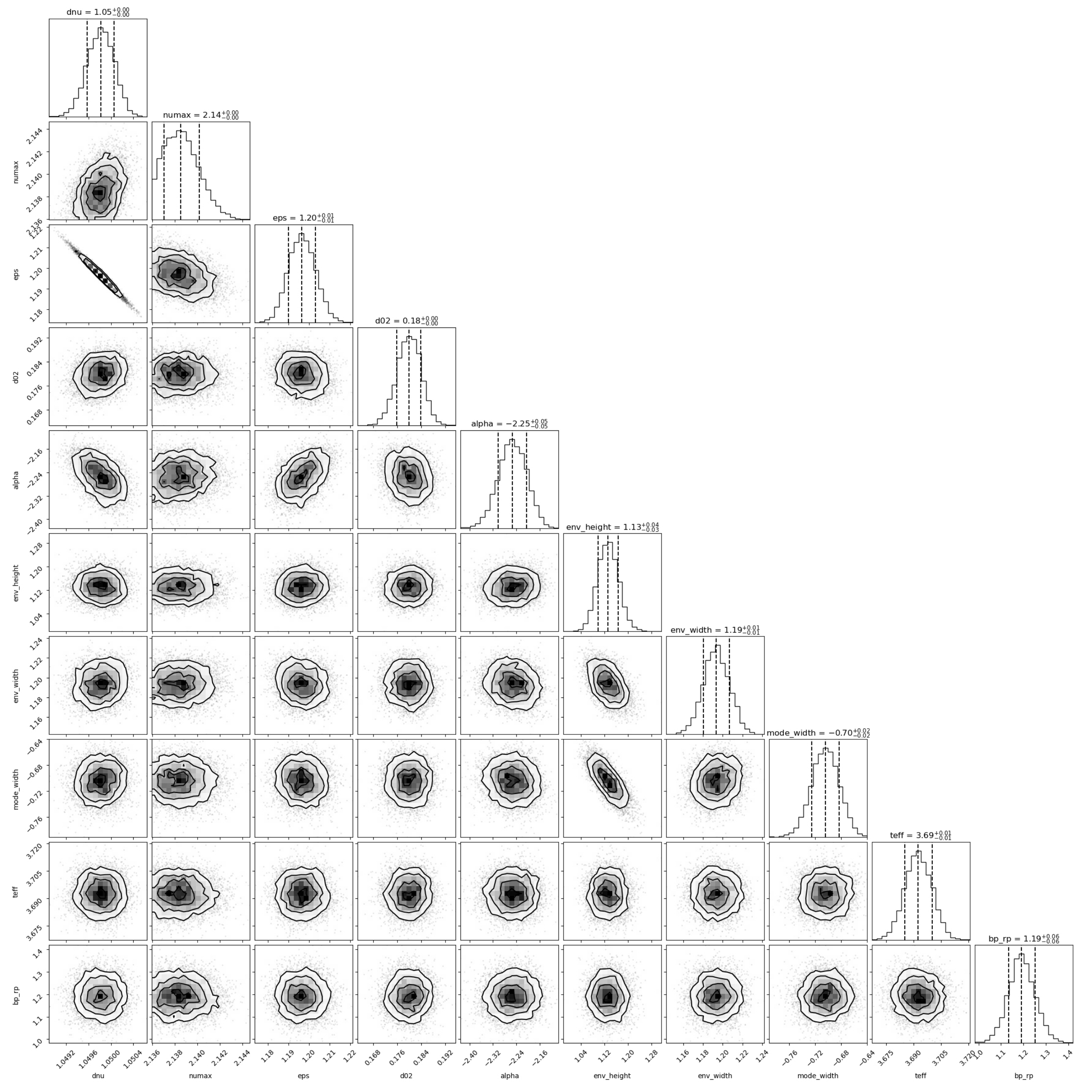}
  \centering
  \caption{HD 33844 corner diagram of asymptotic fit parameters.
  \label{fig:HIP 67851 3}}
  \end{figure}

  \end{document}